\documentclass[acmsmall,screen]{acmart}

\startPage{1}

\setcopyright{rightsretained}
\acmPrice{}
\acmDOI{10.1145/3290388}
\acmYear{2019}
\copyrightyear{2019}
\acmJournal{PACMPL}
\acmVolume{3}
\acmNumber{POPL}
\acmArticle{75}
\acmMonth{1}

\usepackage{tikz}
\usetikzlibrary{positioning}
\usetikzlibrary{matrix}
\usetikzlibrary{fit,calc}

\usepackage{xcolor}
\usepackage{float}
\usepackage{wrapfig}

\usepackage{cleveref}
\crefname{section}{\S}{\S}
\Crefname{Section}{\S}{\S}

\usepackage{commands}
\usepackage{enumitem}
\usepackage{multicol}
\usepackage{enumitem}
\usepackage{hyperref}
\usepackage{subcaption}

\usepackage{listings}
\usepackage{amssymb}

\newsavebox\MBox

\ifdefined\withcolor
	 \definecolor{haskellblue}{rgb}{0.0, 0.0, 1.0}
	 \definecolor{haskellstr}{rgb}{0.2, 0.2, 0.6}
	 \definecolor{haskellred}{rgb}{1.0, 0.0, 0.0}
  \definecolor{gray_ulisses}{gray}{0.55}
  \definecolor{castanho_ulisses}{rgb}{0.71,0.33,0.14}
  \definecolor{preto_ulisses}{rgb}{0.41,0.20,0.04}
  \definecolor{green_ulises}{rgb}{0.2,0.75,0}
\else
	\definecolor{haskellblue}{gray}{0.1}
	\definecolor{haskellstr}{gray}{0.1}
	\definecolor{haskellred}{gray}{0.1}
	\definecolor{gray_ulisses}{gray}{0.1}
	\definecolor{castanho_ulisses}{gray}{0.1}
	\definecolor{preto_ulisses}{gray}{0.1}
	\definecolor{green_ulisses}{gray}{0.1}
\fi

\definecolor{lcolor}{gray}{0.0}
\definecolor{lappcolor}{gray}{0.0}
\definecolor{lappascolor}{gray}{0.0}

\def\codesize{\small}

\definecolor{usercolor}{RGB}{47, 46, 51}
\definecolor{liocolor}{RGB}{10, 55, 104}
\definecolor{yesodcolor}{RGB}{11, 104, 51}
\definecolor{dbcolor}{RGB}{47, 46, 51}

\newcommand\showyesod[1]{\color{yesodcolor}{#1}}
\newcommand\showlio[1]{\color{liocolor}{#1}}

\lstdefinelanguage{HaskellUlisses} {
	basicstyle=\ttfamily\small,
	moredelim=[is][\showyesod]{\*}{\*},
	moredelim=[is][\showlio]{\^}{\^},
	sensitive=true,
	morecomment=[l][\color{gray_ulisses}\ttfamily\itshape\codesize]{--},
	morestring=[b]",
	stringstyle=\color{haskellstr},
	basewidth={0.53em},
	showstringspaces=false,
	numberstyle=\codesize,
	numberblanklines=true,
	showspaces=false,
	breaklines=true,
	showtabs=false,
  literate={ {quals}{{$\mathbb{Q}$}}1
             {iquals}{{$\mathbb{Q}$}}2
             {ltsolzero}{{$A_0$}}2
             {band}{{$\textbf{\texttt{and}}$}}2
             {->}{{$\rightarrow$}}2
             {<-}{{$\leftarrow$}}1
             {monotonic}{{monotonic}}9
             {not}{{$\neg\!\!\!$}}2
             {===}{{$\equiv$}}2
             {ς}{{$\varsigma$}}1
             {ε}{{$\epsilon$}}1
             {φ}{{$\phi$}}1
             {canFlowTo}{{$\sqsubseteq$}}1
             {cannotFlowTo}{{$\not\sqsubseteq$}}1
             {meet}{{$\sqcap$}}1
             {jjoin}{{join}}4
             {join}{{$\sqcup$}}1
             {bot}{{$\perp$}}1
             {top}{{$\top$}}1
             {=>}{{$\Rightarrow$}}1
             {<=>}{{$\Leftrightarrow$}}1
             {GType}{{\tilde{\mathit{\texttt{Type}}}}}4
           },
	emph=
	{[1]
		FilePath,IOError,abs,acos,acosh,all,and,any,appendFile,approxRational,asTypeOf,asin,
		asinh,atan,atan2,atanh,basicIORun,break,catch,ceiling,chr,compare,concat,concatMap,
		const,cos,cosh,curry,cycle,decodeFloat,denominator,digitToInt,div,divMod,drop,
		dropWhile,either,elem,encodeFloat,enumFrom,enumFromThen,enumFromThenTo,enumFromTo,
		error,even,exp,exponent,fail,mapMaybe,filter,flip,floatDigits,floatRadix,floatRange,floor,
		fmap,foldl,foldl1,foldr,foldr1,fromDouble,fromEnum,fromInt,fromInteger,fromIntegral,
		fromRational,fst,gcd,getChar,getContents,getLine,head,id,inRange,index,init,intToDigit,
		interact,ioError,isAlpha,isAlphaNum,isAscii,isControl,isDenormalized,isDigit,isHexDigit,
		isIEEE,isInfinite,isLower,isNaN,isNegativeZero,isOctDigit,isPrint,isSpace,isUpper,iterate,
		last,lcm,length,lex,lexDigits,lexLitChar,lines,log,logBase,lookup,map,mapM,mapM_,max,
		maxBound,posMax,negMax,maximum,maybe,min,minBound,minimum,mod,negate,not,notElem,null,numerator,odd,
		or,ord,pi,pred,primExitWith,print,product,properFraction,putChar,putStr,putStrLn,quot,
		quotRem,range,rangeSize,read,readDec,readFile,readFloat,readHex,readIO,readInt,readList,readLitChar,
		readLn,readOct,readParen,readSigned,reads,readsPrec,realToFrac,recip,rem,repeat,replicate,return,
		round,scaleFloat,scanl,scanl1,scanr,scanr1,seq,sequence,sequence_,show,showChar,showInt,
		showList,showLitChar,showParen,showSigned,showString,shows,showsPrec,significand,signum,sin,
		sinh,snd,span,splitAt,sqrt,subtract,succ,sum,tail,take,takeWhile,tan,tanh,threadToIOResult,toEnum,
		toInt,toInteger,toLower,toRational,toUpper,truncate,uncurry,undefined,unlines,until,unwords,unzip,
		unzip3,userError,words,writeFile,zip,zip3,zipWith,zipWith3,listArray,doParse,empty,for,initTo,
        assert,compose,checkGE,maxEvens,empty,create,get,set,initialize,idVec,fastFib,fibMemo,
        ex1,ex2,ex3,incr,inc,dec,isPos,positives,find,insert,len,size,union,fromList,initUpto,trim,
        insertSort,decsort,qsort,reverse,append,upperCase, ifM, whileM, get, decrM, diff,
        project, select, leq, elts, keys, dkeys, dfun, addKey, pTrue, emptyRD, rFalse,
        	dom, rng, isI, isD, isS, movie1, movie2,  toI, toS, toD, good_titles, runState, ret,
        	update, getCtr, setCtr, ctr, rdCtr, wrCtr, ifTest, whileTest, posCtr, zeroCtr, decr, decCtr,
        	pread , pwrite , plookup , pcontents, pcreateF , pcreateFP, pcreateD, active, caps, pset, eqP,
        	write, contents, alloc, derivP, copyP, createDir, store, copyRec, copySpec,
        	forM_, when, flookup, fread, createDir, pcreateFile, isFile, copyFrame, ?
	},
	emphstyle={[1]\color{haskellblue}},
	emph=
	{[2] Eq, Program, Label, 
	},
	emphstyle={[2]\color{castanho_ulisses}},
	emph=
	{[3]
		case,class,data,deriving,do,else,if,import,in,infixl,infixr,instance,let,
		module,of,primitive,then,refinement,type,where,forall,bound,and,
		measure,reflect,predicate, assume, return
	},
	emphstyle={[3]\color{preto_ulisses}\textbf},
	emph=
	{[4]
		quot,rem,div,mod,elem,notElem,seq
	},
	emphstyle={[4]\color{castanho_ulisses}\textbf},
	emph=
	{[5]
		EQ,GT,LT,Left,Right
	},
	emphstyle={[5]\color{preto_ulisses}\textbf},
	emph=
	{[6]
	    axiomatize, measure, inline, return
	},
	emphstyle={[6]\color{lcolor}}
}

\lstnewenvironment{code}
{\lstset{language=HaskellUlisses}}
{}

\lstnewenvironment{mcode}
{\lstset{language=HaskellUlisses,columns=fullflexible,keepspaces,mathescape}}
{}

\lstMakeShortInline[language=HaskellUlisses,mathescape,keepspaces,mathescape,basicstyle=\ttfamily\small,breakatwhitespace]@

\lstdefinelanguage{Pseudo} {
	basicstyle=\ttfamily\codesize,
	sensitive=true,
  mathescape=true,
	morecomment=[l][\color{gray_ulisses}\ttfamily\codesize]{--},
	morecomment=[s][\color{gray_ulisses}\ttfamily\codesize]{\{-}{-\}},
	morestring=[b]",
	showstringspaces=false,
	numberstyle=\codesize,
	numberblanklines=true,
	showspaces=false,
	breaklines=true,
	showtabs=false
}

\begin{document}


\title[LWeb: Information Flow Security for Multi-tier Web Applications]{LWeb: Information Flow Security for Multi-tier Web Applications}

\author{James Parker}
\affiliation{
  \department{Department of Computer Science}              
  \institution{University of Maryland}            
  \country{USA}                    
}

\author{Niki Vazou}\authornote{Vazou was at the University of Maryland
  while this work was carried out.}
 \affiliation{
   \institution{IMDEA Software Institute}            
   \country{Spain}                    
}

\author{Michael Hicks}
\affiliation{
  \department{Department of Computer Science}              
  \institution{University of Maryland}            
  \country{USA}                    
 }

\begin{abstract}
  This paper presents \lweb, a framework for enforcing label-based,
  information flow policies in database-using web applications. In a
  nutshell, \lweb marries the \lio Haskell IFC enforcement library
  with the \yesod web programming framework. The implementation has
  two parts. First, we extract the core of \lio into a monad
  transformer (\lmonad) and then apply it to \yesod's core monad. Second,
  we extend \yesod's table definition DSL and query functionality to
  permit defining and enforcing label-based policies on tables and enforcing
  them during query processing. \lweb's policy language is expressive,
  permitting dynamic per-table and per-row policies. We formalize the
  essence of \lweb in the \lwebcalc calculus and mechanize the proof of
  noninterference in Liquid Haskell. This mechanization constitutes
  the first metatheoretic proof carried out in Liquid Haskell. We also
  used \lweb to build a substantial web site hosting the \emph{Build it,
  Break it, Fix it} security-oriented programming contest. The site
  involves 40 data tables and sophisticated policies. Compared to
  manually checking 
  security policies, \lweb imposes a modest runtime overhead of between \overheadnumbermin to
  \overheadnumber. It reduces the trusted code base from the
  whole application to just \tcbnumberbibifi of the application code, and
  \tcbnumber of the code overall (when counting \lweb too).
\end{abstract}





\maketitle

\section{Introduction}\label{sec:intro}

Modern web applications must protect the confidentiality and integrity
of their data. Employing
access control and/or manual, ad hoc enforcement mechanisms may fail to block illicit
information flows between components, \eg from database to server to
client. 
Information flow control (IFC)~\cite{sabelfeld:survey} policies can
govern such flows, but enforcing them poses practical problems.
Static enforcement (\eg by
typing~\cite{jif,flowcaml,Chong:2007:SEC:1362903.1362904,Schoepe:2014:STI:2628136.2628151,Chong:2007:SWA:1294261.1294265}
or static
analysis~\cite{Hammer:2009:FCO:1667545.1667547,JohnsonWMC2015,Arzt:2014:FPC:2666356.2594299})
can produce too many false
alarms, which hamper adoption~\cite{king08implicit}. Dynamic
enforcement~\cite{Chudnov:2015:IIF:2810103.2813684,Roy:2009:LPF:1542476.1542484,Tromer:2016:DII:2897845.2897888,YangHASFC16,Austin:2012:MFD:2103656.2103677}
is more precise
but can impose high overheads.

A promising solution to these problems is embodied in the LIO
system~\cite{lio} for Haskell. LIO is a drop-in replacement for the
Haskell IO monad, extending IO with an internal \emph{current label}
and \emph{clearance label}. Such labels are lattice ordered (as is
typical~\cite{denning}), with the degenerate case being a secret (high)
label and public (low) one. LIO's current label constitutes the least
upper bound of the security labels of all values read during the
current computation. Effectful operations such as reading/writing from
stable storage, or communicating with other processes, are checked
against the current label. If the operation's security label (\eg
that on a channel being written to) is lower than the current label,
then the operation is rejected as potentially insecure. The clearance
serves as an upper bound that the current label may never cross, even
prior to performing any I/O, so as to reduce the chance of side
channels. Haskell's clear, type-enforced separation of pure
computation from effects makes LIO easy to implement soundly and
efficiently, compared to other dynamic enforcement mechanisms. 

This paper presents \lweb, an extension to \lio that aims to bring its
benefits to Haskell-based web applications. This paper presents the
three main contributions of our work. 

First, we present an extension to a core LIO formalism with support
for database transactions. Each table has a label that protects its
length. In our implementation we use DC labels~\cite{stefan:dclabels},
which have both confidentiality and integrity components. 
The confidentiality component of the table label controls who can
query it (as the result may reveal something about the table's
length), and the integrity component controls who can add or delete
rows (since both may change the length). In addition, each row may
have a more refined policy to protect its contents. The label for a
field in a row may be specified as a function of other fields in the
same row (those fields are protected by a specific, global
label). This allows, for example, 
having a row specifying a
user and some sensitive user data; the former can act as a label to
protect the latter. 

We mechanized our formalism in Liquid
Haskell~\cite{Vazou14} and proved that it enjoys noninterference. Our
development proceeds in two steps: a core \lio formalism
called \liocalc (\cref{sec:formal}), and an extension to it,
called \lwebcalc, that adds database operations
(\cref{sec:formal-db}). The mechanization 
process was fruitful: it revealed two bugs in our original rules
that constituted real 
leaks. Moreover, this mechanization constitutes the
largest-ever development in Liquid Haskell
and is the first Liquid Haskell application to prove a 
language 
metatheory
(\cref{sec:liquidhaskell-discussion}). 

As our next contribution, we describe a full implementation of 
\lweb in Haskell as an
extension to the \yesod web programming framework
(\cref{sec:overview} and~\cref{sec:impl}). Our implementation
was carried out in two steps. First, we extracted the core label
tracking functionality of \lio into a monad transformer called
\lmonad so that it can be layered on monads other than @IO@\@. For \lweb, we
layered it on top of the @Handler@ monad provided by the \yesod. This
monad encapsulates mechanisms for client/server HTTP communications
and database transactions, so layering \lmonad on top of @Handler@
provides the basic functionality to enforce security. Then we extended 
\yesod's database API to permit defining label-based information flow
policies, generalizing the approach from our formalism whereby each
row may have many fields, each of which may be protected by other
fields in the same row. We support simple key/value lookups and more
general SQL queries, extending the Esqueleto framework~\cite{esqueleto}. We use Template
Haskell~\cite{Sheard:2002:TMH:636517.636528} to insert checks that properly enforce policies in our
extension.

Finally, we describe our experience using \lweb to
build a substantial web site hosting the Build it,
Break it, Fix it (\bibifi) security-oriented programming
contest~\cite{Ruef:2016:BBF:2976749.2978382} hosted at
\url{https://builditbreakit.org} (\cref{subsec:bififi}). This site
has been used over the last few years to host more than a dozen
contests involving hundreds of teams. It consists of 11500+ lines of
Haskell and manages data stored in 40 database tables. The site has a
variety of roles (participants, teams, judges, admins) and policies
that govern their various privileges. When we first deployed this
contest, it lacked \lweb support, and we found it had 
authorization bugs. Retrofitting it with \lweb was straightforward
and eliminated those problems, reducing the trusted computing base
from the entire application to just 80 lines of its code
(\tcbnumberbibifi) plus the \lweb codebase (for a total of
\tcbnumber). \lweb imposes modest overhead on \bibifi 
query latencies---experiments show between \overheadnumbermin and
\overheadnumber (\cref{sec:experiments}). 

\lweb is not the first framework to use IFC to enforce database
security in web applications. Examples of prior efforts include
SIF/Swift~\cite{Chong:2007:SEC:1362903.1362904,
  Chong:2007:SWA:1294261.1294265}, Jacqueline~\cite{YangHASFC16},
Hails~\cite{Giffin:2012:HPD:2387880.2387886,stefan17hails},
SELinks~\cite{corcoran09selinks},
SeLINQ~\cite{Schoepe:2014:STI:2628136.2628151}, 
UrFlow~\cite{urflow}, and
IFDB~\cite{Schultz:2013:IDI:2465351.2465357}. \lweb distinguishes
itself by providing end-to-end IFC security (between/across server and database), backed
by a formal proof (mechanized in Liquid Haskell), for a mature,
full-featured web framework (\yesod) while supporting expressive
policies (\eg where one field can serve as the label of another) and
efficient queries (a large subset of SQL). The IFC checks needed
during query processing were tricky to get right---our formalization
effort uncovered bugs in our original implementation by which
information could leak owing to the checks
themselves. \Cref{sec:related} discusses related work in detail. 

The code for \lweb and its mechanized proof are freely available.

\section{Overview}\label{sec:overview}
%
\begin{wrapfigure}{r}{0.3\textwidth}
\vspace{-1.8cm}
\begin{tikzpicture}[
mynode/.style={
  draw,
  text width=3cm,
  minimum height=0.7cm,
  align=center
  },
]
\node[mynode,fill=usercolor] (user) {\textcolor{white}{\textbf{Programmer}}};

\node[below=1cm of user] (lwebname) {$\qquad\qquad\qquad\qquad$ \textbf{\lweb}};
\node[mynode,fill=liocolor,below=0.2cm of lwebname] (lio) {\textcolor{white}{\textbf{\lmonad}}};
\node[mynode,fill=yesodcolor,below=1cm of lio] (yesod) {\textcolor{white}{\textbf{\yesod}}};

\node [draw=black,minimum width=4cm, yshift=0.5cm, fit={ (lio) (yesod) (lwebname)}, below=1.5cm of user] (lweb){};

\node[mynode,fill=dbcolor,below=1cm of lweb] (db) {\textcolor{white}{\textbf{DB}}};

\draw[<->,very thick] 
  (user) -- node[left] {DB Query} (lweb);
\draw[<->,very thick] 
  (lio) -- node[left] {Label Check} (yesod);
\draw[<->,very thick] 
  (db) -- node[left] {DB Access} (lweb);
\end{tikzpicture}
\caption{Structure of \lweb.}
\label{fig:structure}
\vspace{1.0cm}
\end{wrapfigure}
%
The architecture of \lweb is shown
in~\cref{fig:structure}.
Database queries/updates precipitated by user interactions are
processed by the \lmonad component, which constitutes the core of \lio 
and confirms that label-based
security policies are not violated. Then, the queries/updates are
handled via \yesod, where the results continue to be subject to policy
enforcement by \lmonad.
\begin{wrapfigure}{r}{.3\textwidth}
\vspace{-1.6cm}
\centering
\begin{mcode}
  class Eq a => Label a where
    bot   :: a 
    (join) :: a -> a -> a 
    (meet) :: a -> a -> a
    (canFlowTo) :: a -> a -> Bool
\end{mcode}
\caption{The \texttt{Label} class}
\label{fig:label}
\end{wrapfigure}


\subsection{Label-Based Information Flow Control with \lio}
\label{sec:lio-intro}

We start by presenting \lio~\cite{lio} and how it is used 
to enforce noninterference for label-based information flow policies.

\paragraph{Labels and noninterference}
As a trivial security label, consider a datatype with constructors
@Secret@ and @Public@.
Protected data is assigned a label, and an IFC system ensures that 
@Secret@-labeled data can only be learned by those with
@Secret@-label privilege or greater.
The label system can be generalized to any 
lattice~\cite{denning} where IFC is checked using the lattice's 
partial order relation @canFlowTo@. Such a system enjoys
\emph{noninterference}~\cite{goguen} if an adversary with privileges
at label @l1@ can learn nothing about data labeled with @l2@ where
@l2@ $\not\sqsubseteq$ @l1@.

In~\cref{fig:label}
we define the label interface as the type class @Label@
that defines the bottom (least protected) label, least upper bound
(join, $\sqcup$) of two labels, the greatest lower bound (meet, $\sqcap$), 
and whether one label can flow to ($\sqsubseteq$) another, defining a
partial ordering. Instantiating this type class for
@Public@ and @Secret@ would set @Public@ as the bottom label and
@Public@ $\sqsubset$ @Secret@ (with join and meet operations to match).

\paragraph{The LIO monad}
\lio enforces IFC on labeled data using dynamic checks. The type
@LIO l a@ denotes a monadic computation that returns a value of type
@a@ at label @l@. \lio provides two methods to label and unlabel data.  
\begin{code}
  label   :: (Label l) => l -> a -> LIO l (Labeled l a)
  unlabel :: (Label l) => Labeled l a -> LIO l a
\end{code} 
The method @label l v@ takes as input a label and some data and
returns a @Labeled@ value, \ie the data @v@ marked with the label @l@. 
The method @unlabel v@ takes as input a labeled value 
and returns just its data. The \lio monad maintains an ambient
label---the \emph{current label} @lc@---that represents the label of the
current computation. As such, labelling and unlabelling a
value affects @lc@. In particular, @unlabel v@ updates
@lc@ by joining it to @v@'s label, while @label l v@ is
only permitted if @lc@ $\sqsubseteq$ @l@, \ie the current label can
flow to @l@. If this check fails, \lio raises an exception. 

As an example, on the left, a computation with current label @Public@
labels data @"a secret"@ as @Secret@, preserving the same current label, 
and then unlabels the data, thus raising the current label to @Secret@.
On the right, a computation with current label @Secret@ 
attempts to label data as @Public@, which fails, 
since the computation is already tainted with (\ie dependent on) secret data. 
\begin{flushleft}
\begin{tabular}{lcl}
\begin{mcode} 
  -- lc := Public
  v <- label Secret "a secret"
  -- ok: Public canFlowTo Secret and lc := Public 
  x <- unlabel v
  -- lc := Secret 
\end{mcode}
&\quad\quad\quad&
\begin{mcode} 
  -- lc := Secret
  v <- label Public "public"
  -- exception: Secret cannotFlowTo Public
  
  $\quad$ 
\end{mcode}
\end{tabular}
\end{flushleft}

\lio also supports labeled mutable references, and a scoping mechanism for
temporarily (but safely) raising the current label until a
computation completes, and then restoring it. \lio also has what
is called the \emph{clearance} label that serves as an upper bound 
for the current label, and thus can serve to identify potentially
unsafe computations sooner.

A normal Haskell program can run an \lio computation via %
@runLIO@, whose type is as follows. 
\begin{mcode}
  runLIO :: (Label l) => LIO l a -> IO a
\end{mcode}
Evaluating @runLIO m@ initializes the current label to $\bot$ and
computes @m@. The returned result is an @IO@ computation, since \lio
allows @IO@ interactions, \eg with a file system. If any security
checks fail, @runLIO@ throws an exception.


\subsection{\yesod}
\label{sec:yesod}

\yesod~\cite{yesod} is mature framework for developing type-safe  
and high performance web applications in Haskell. In a nutshell, \lweb
adds \lio-style support to \yesod-based web applications, with a focus on
supporting database security policies.

\begin{figure}
\centering
\begin{minipage}{.7\textwidth}
\begin{mcode}
  *Friends* ^<bot,Const Admin>^
   *user1 Text* ^<bot,Const Admin>^
   *user2 Text* ^<bot,Const Admin>^
   *date  Text* ^<Field User1 meet Field User2,Const Admin>^
\end{mcode}
\end{minipage}
\caption{Example \lweb database table definition.
The \textcolor{yesodcolor}{green} is \texttt{Yesod} syntax
and the \textcolor{liocolor}{blue} is the \textit{LWeb} policy.} 
\label{fig:friendstable}
\end{figure}
The \textcolor{yesodcolor}{green} part of \cref{fig:friendstable} uses 
\yesod's domain specific language (DSL) to define the table @Friends@. 
The table has three @Text@\footnote{\texttt{Text} is an efficient Haskell string type.} 
fields corresponding to two users (@user1@ and @user2@) and the date of their friendship. 
A primary key field with type @FriendsId@ is also automatically added.
In~\cref{subsec:lweb} we explain how the \textcolor{liocolor}{blue} part of the definition 
is used for policy enforcement. 

\yesod uses Template Haskell~\cite{Sheard:2002:TMH:636517.636528}
to generate, at compile time, a database schema from such table definitions. 
These are the Haskell types that \yesod generates for the @Friends@ table. 
\begin{code}
  data FriendsId = FriendsId Int
  data Friends = Friends { friendsUser1 :: Text, friendsUser2 :: Text
                         , friendsDate  :: Text }
\end{code}
Note that though each row has a key of type @FriendsId@, it is elided
from the @Friends@ data record. Each generated key type is a member of
the @Key@ type family; in this case @Key Friends@ is a type alias for
@FriendsId@.

\yesod provides an API to define and run queries. 
Here is a simplified version of this API. 
\begin{mcode}
  runDB  :: YesodDB a -> Handler a 
  get    :: Key v -> YesodDB (Maybe v)
  insert :: v     -> YesodDB (Key v)
  delete :: Key v -> YesodDB ()
  update :: Key v -> [Update v] -> YesodDB ()
\end{mcode}
The type alias @YesodDB a@ denotes 
the monadic type of a computation that queries (or updates) the database. 
The function @runDB@ runs the query argument on the database.
@Handler@ is \yesod's underlying monad used to respond to HTTP requests. 
The functions @get@, @insert@, @delete@, and @update@
generate query computations.
For example, we can query the database for the date of a specific friendship using @get@. 
\begin{code}
  getFriendshipDate :: FriendsId -> Handler (Maybe Text)
  getFriendshipDate friendId = do
      r <- runDB (get friendId)
      return (friendsDate <$> r)
\end{code}


\yesod also supports more sophisticated SQL-style queries via an interface
called Esqueleto~\cite{esqueleto}. Such queries may include inner and outer
joins, conditionals, and filtering.

\subsection{\lweb: \yesod with \lio}\label{subsec:lweb}

\lweb extends \yesod with \lio-style IFC enforcement. The
implementation has two parts. As a first step, we generalize 
\lio to support an arbitrary underlying monad by making it a
\emph{monad transformer}, applying it to \yesod's core monad.
Then we extend \yesod operations to incorporate label-based
policies that work with this extended monad. 

\paragraph{\lmonad: LIO as a monad transformer}
\lmonad generalizes the underlying @IO@ monad of \lio to \textit{any}
monad @m@. In particular, \lmonad is a
monad transformer @LMonadT l m@ that adds the 
IFC operations to the underlying monad @m@, rather than making it
specific to the @IO@ monad.
\begin{mcode}
  label     :: (Label l, Monad m) => l -> a -> LMonadT l m (Labeled l a)
  unlabel   :: (Label l, Monad m) => Labeled l a -> LMonadT l m a
  runLMonad :: (Label l, Monad m) => LMonadT l m a -> m a
\end{mcode}
@LMonadT@ is implemented as a state monad transformer that tracks the current label. 
Computations that run in the underlying @m@ monad cannot be executed directly due to Haskell's type system. 
Instead, safe variants that enforce IFC must be written so that they can be executed in @LMonadT l m@. 
Thus, the \lio monad is an instantiation of 
the monad variable @m@ with @IO@: @LIO l = LMonadT l IO@.
For \lweb we instantiate @LMonadT@ with \yesod's @Handler@ monad.
\begin{mcode}
  type LHandler l a = LMonadT l Handler a
\end{mcode}
Doing this adds information flow checking to \yesod applications, but
it still remains to define policies to be checked. Thus we extend \yesod
to permit defining label-based policies on database schemas, and to
enforce those policies during query processing.

\paragraph{Label-annotated database schemas}
\lweb labels are based on DC labels~\cite{stefan:dclabels}, which have
the form @<l,r>@, where the left protects the \emph{confidentiality} and the right
protects the \emph{integrity} of the labeled value. Integrity lattices are
dual to confidentiality lattices. They track who can influence the
construction of a value. 

Database policies are written as label annotations @p@ on table
definitions, following this grammar: 
\begin{mcode}
  p := <l, l>
  l := Const c | Field f | Id | top | bot | l meet l | l join l 
\end{mcode}
Here, @c@ is the name of a data constructor and @f@ is a field
name. A database policy consists of a single \emph{table label} and one
label for each field in the database. We explain these by example.

The security labels of the @Friends@ table are given by the
\textcolor{liocolor}{blue} part of \cref{fig:friendstable}.
The first line's label @*Friends* ^<bot,Const Admin>^@
defines the table label, which protects the \emph{length} of
the table. This example states that anyone can learn the length of the
table (\eg by querying it), but only the administrator can change
the length (\ie by adding or removing entries). 
\lweb requires the table label to be constant, \ie 
it may not depend on run-time entries of the table. Allowing it to do so
would significantly complicate enforcing noninterference.

The last line @*date Text* ^<Field User1 meet Field User2,Const Admin>^@
defines that either of the users listed in the first two fields can read the
@date@ field but only the administrator can write it.  
This label is \emph{dynamic}, since the values of the @user1@ and @user2@
fields may differ from row to row. 
We call fields, like @user1@ and @user2@, which are referenced in
another field's label annotation, \emph{dependency fields}. When a
field's label is not given explicitly, the
label @<bot,top>@ is assumed. To simplify security enforcement, \lweb
requires the label of a dependency field to be constant and flow into
(be bounded by) the table label. For @user1@ and @user2@ this holds
since their labels match the table's label. 

The invariants about the table label and the dependency field labels
are enforced by a compile-time check, when processing
the table's policy annotations. Note that @Labeled@ values may not be
directly stored in the database as there is no way to directly express
such a type in a source program. Per \cref{fig:friendstable}, field types
like @Text@, @Bool@, and @Int@ are allowed, and their effective label
is indicated by annotation, rather than directly expressed in the
type.\footnote{The formalism encodes all of these invariants with
  refinement types in the database definition.} 
%
%

\paragraph{Policy enforcement}
\lweb enforces the table-declared policies by providing wrappers around each \yesod 
database API function.
\begin{mcode}
  runDB  :: Label l => LWebDB l a -> LHandler l a 
  get    :: Label l => Key v -> LWebDB l (Maybe v)
  insert :: Label l => v     -> LWebDB l (Key v)
  delete :: Label l => Key v -> LWebDB l ()
  update :: Label l => Key v -> [Update v] -> LWebDB l ()
\end{mcode}
Now the queries are modified to return @LWebDB@ computations 
that are evaluated (using @runDB@) inside the @LHandler@ monad. 
For each query operation, 
\lweb wraps the underlying database query with information flow control checks that enforce the defined policies. 
For instance, if @x@ has type @FriendsId@, 
then @r <- runDB $\$$ get x@ joins the current label 
with the label of the selected row, here @user1 meet user2@.
%

\lweb also extends IFC checking to advanced SQL queries expressed in
Esqueleto~\cite{esqueleto}. As explained in~\cref{sec:impl}, \lweb uses a DSL syntax,
as a @lsql@ quasiquotation, to wrap these queries with IFC checks. 
For example, the following query joins the @Friends@ table with a @User@ table: 

\begin{mcode}
  rs <- runDB [lsql|select $\star$ from Friends inner jjoin User on Friends.user1 == User.id|]
\end{mcode}

%
%

\section{Mechanizing Noninterference of LIO in Liquid Haskell}\label{sec:formal}

A contribution of this work is a formalization of \lweb's extension to
\lio to support database security policies, along with a proof that
this extension satisfies (termination insensitive) noninterference. We
mechanize our formalization in Liquid Haskell~\cite{Vazou14}, an
SMT-based refinement type checker for Haskell programs. Liquid Haskell
permits refinement type specifications on Haskell source code. It
converts the code into SMT queries to validate that the code satisfies
the specifications. Our mechanized formalization and proof of
noninterference constitutes the first significant metatheoretical
mechanization carried out in Liquid Haskell.

We present our mechanized \lweb formalism in two parts. In this
section, we present \liocalc, a formalization and proof of
noninterference for \lio. The next section presents \lwebcalc, an
extension of \liocalc that supports database operations. Our Liquid
Haskell mechanization defines \liocalc's syntax and operational
semantics as Haskell definitions, as a 
definitional interpreter. We present them the same way in this paper,
rather than reformatting them as mathematical inference rules.
Metatheoretic properties are expressed
as refinement types, following~\citet{refinement-reflection,a-tale},
and proofs are Haskell functions with these types (checked by
the SMT solver). We assess our experience using Liquid Haskell for
metatheory in comparison to related approaches in
\Cref{sec:liquidhaskell-discussion}.  

\subsection{Security Lattice as a Type Class}\label{subsec:label:class}
\begin{figure*}
\begin{mcode}
  class Label l where
    (canFlowTo) :: l -> l -> Bool
    (meet) :: l -> l -> l
    (join) :: l -> l -> l
    bot   :: l

    lawBot              :: l:l  -> { bot canFlowTo l }
    lawFlowReflexivity  :: l:l  -> { l canFlowTo l }
    lawFlowAntisymmetry :: l1:l -> l2:l -> { (l1 canFlowTo l2 $\land$ l2 canFlowTo l1) => l1 == l2 }
    lawFlowTransitivity :: l1:l -> l2:l -> l3:l -> { (l1 canFlowTo l2 $\land$ l2 canFlowTo l3) => l1 canFlowTo l3 }

    lawMeet :: z:l -> l1:l -> l2:l -> l:l 
            -> { z == l1 meet l2 => z canFlowTo l1 $\land$ z canFlowTo l2 $\land$ (l canFlowTo l1 $\land$ l canFlowTo l2 => l canFlowTo z) }
    lawJoin :: z:l -> l1:l -> l2:l -> l:l 
            -> { z == l1 join l2 => l1 canFlowTo z $\land$ l2 canFlowTo z $\land$ (l1 canFlowTo l $\land$ l2 canFlowTo l => z canFlowTo l) }

\end{mcode}
\caption{\texttt{Label} type class extended with \texttt{law*} methods to
  define the lattice laws as refinement types.}
\label{fig:formalism:label}
\end{figure*}
Figure~\ref{fig:formalism:label} duplicates the @Label@ class
definition of Figure~\ref{fig:label} but extends it with several
methods that use refinement types to express properties of
lattices that labels are expected to have.

\paragraph{Partial order}
The method @(canFlowTo)@ defines a partial order for each @Label@ element. 
That is, @(canFlowTo)@ is reflexive, antisymmetric, and transitive, 
as respectively encoded by the refinement types of the methods 
@lawFlowReflexivity@, @lawFlowAntisymmetry@, and @lawFlowTransitivity@. 
For instance, 
@lawFlowReflexivity@ is a method that takes a label 
@l@ to a Haskell unit (\ie @l -> ()@).
This type is refined to encode the reflexivity property
@l:l  -> {v:() | l canFlowTo l }@
and further simplifies to ignore the irrelevant @v:()@ part as 
@l:l  -> { l canFlowTo l }@.
With that refinement, application of @lawFlowReflexivity@ to a concrete label @l@
gives back a proof that @l@ can flow to itself (\ie @l canFlowTo l@).
At an instance definition of the class @Label@,
the reflexivity proof needs to be explicitly provided. 

\paragraph{Lattice}
Similarly, we refine the @lawMeet@ method to define the properties of the @(meet)@ lattice operator. 
Namely, for all labels @l1@ and @l2@, we define @z == l1 meet l2@ so that 
(i) @z@ can flow to @l1@ and @l2@ (@z canFlowTo l1 $\land$ z canFlowTo l2@) and 
(ii) all labels that can flow to @l1@ and @l2@, can also flow to @z@ 
@(forall l. l canFlowTo l1 $\land$ l canFlowTo l2 => l canFlowTo z)@.
Dually, we refine the @lawJoin@ method 
to describe @l1 join l2@ as the minimum label 
that is greater than @l1@ and @l2@. 

\paragraph{Using the lattice laws}
The lattice laws are class methods, which can be used for each @l@ that satisfies the 
@Label@ class constraints. 
For example, we prove that for all labels @l1@, @l2@, and @l3@, 
@l1 join l2@ cannot flow into @l3@ \textit{iff}
@l1@ and @l2@ cannot both flow into @l3@.
\begin{mcode}
  $\texttt{join}$Iff :: Label l => l1:l -> l2:l -> l3:l -> {l1 canFlowTo l3 $\land$ l2 canFlowTo l3 <=> (l1 join l2) canFlowTo l3}
  $\texttt{join}$Iff l1 l2 l3 = lawJoin (l1 join l2) l1 l2 l3 ? lawFlowTransitivity l1 l2 l3 
\end{mcode}
The theorem is expressed as a Haskell function that
is given three labels and returns a unit value 
refined with the desired property.
The proof proceeds by calling the laws of join and transitivity, combined 
with the proof combinator @(?)@ that ignores its second argument 
(\ie defined as @x ? _ = x@)
while passing the refinements of both arguments to the SMT solver.
The contrapositive step is automatically enforced by refinement type checking, using the SMT solver.

\subsection{\liocalc: Syntax and Semantics}\label{subsec:label:calculus}

Now we present the syntax and operational semantics of \liocalc.
\begin{figure}[t]
\begin{tabular}{l}
\begin{mcode}
data Program l = Pg { pLabel :: l, pTerm :: Term l } | PgHole 

data Term l 
 -- pure terms 
 = TUnit | TInt Int | TLabel l | TLabeled l (Term l) | TLabelOf (Term l) 
 | TVar Var | TLam Var (Term l) | TApp (Term l) (Term l) | THole | ...   
 -- monadic terms 
 | TBind   (Term l) (Term l) | TReturn  (Term l) | TGetLabel | TLIO (Term l) 
 | TTLabel (Term l) (Term l) | TUnlabel (Term l) | TException 
 | TToLabeled (Term l) (Term l) 
\end{mcode}
\end{tabular}
\caption{Syntax of \liocalc.}
\label{fig:label:syntax}
\end{figure}

\subsubsection{Syntax}
\Cref{fig:label:syntax} defines a program as either an actual program (@Pg@)
with a current label @pLabel@
under which the program's term @pTerm@ is evaluated, 
or as a hole (@PgHole@).
The hole is not a proper program; it is used for to define adversary
observability when proving noninterference
(\cref{subsec:formal:noninterference}). 
We omit the clearance label in the formalism as a simplification since 
its rules are straightforward (when the current label changes, check that it flows into the clearance label).
%
Terms are divided into 
\textit{pure} terms whose evaluation is independent of the current
label and 
\textit{monadic} terms,
which either manipulate or whose evaluation depends on the current label.

\paragraph{Pure terms}
Pure terms include 
unit @TUnit@, integers @TInt i@ for some Haskell integer @i@, and the label value @TLabel l@, 
where @l@ is some instance of the labeled class of \Cref{fig:formalism:label}.
The labeled value @TLabeled l t@ wraps the term @t@ with the label @l@.
The term @TLabelOf t@ returns the label of the term @t@, if @t@ is a labeled term. 
Pure terms include the standard lambda calculus terms for 
variables (@TVar@), application (@TApp@) and abstraction (@TLam@).
%
Finally, similar to programs, a hole term (@THole@) is required for the meta-theory.
It is straightforward to extend pure terms to more interesting calculi. 
In our mechanization we extended pure terms with 
lattice label operations, 
branches, lists, and inductive fixpoints; we omit them here for space
reasons. 

\paragraph{Monadic terms}
Monadic terms are evaluated under 
a state that captures the current label.
Bind (@TBind@) and return (@TReturn@) are the standard monadic operations, 
that respectively propagate and return the current state.  
The current label is accessed with the @TGetLabel@ term
and the monadic term @TLIO@ wraps monadic values, \ie computations
that cannot be further evaluated. 
The term @TTLabel lt t@ labels the term @t@ with the label term @lt@ and dually 
the term @TUnlabel t@ unlabels the labeled term @t@. 
An exception (@TException@) is thrown if a policy is violated. 
Finally, the term @TToLabeled tl t@  locally raises the current label to @tl@
to evaluate the monadic term @t@, dropping it again when the computation
completes. 

\subsubsection{Semantics}
\begin{figure}[t]
\begin{tabular}{lcl}
\begin{mcode}
eval :: Label l => Program l -> Program l
\end{mcode}\hspace{-2em} &&\\
\begin{mcode}
eval (Pg lc (TBind t1 t2))
 | Pg lc' (TLIO t1') <- eval$*$ (Pg lc t1)
 = Pg lc' (TApp t2 t1')
 
eval (Pg lc (TReturn t)) 
  = Pg lc (TLIO t)

eval (Pg lc TGetLabel)
  = Pg lc (TReturn (TLabel lc))

eval (Pg lc (TTLabel (TLabel l) t))
  | lc canFlowTo l 
  = Pg lc (TReturn (TLabeled l t))
  | otherwise 
  = Pg lc TException
\end{mcode} &&
\begin{mcode}
eval (Pg lc (TUnlabel (TLabeled l t)))
  = Pg (l join lc) (TReturn t)

eval (Pg lc (TToLabeled (TLabel l) t))
  | Pg lc' (TLIO t') <- eval$*$ (Pg lc t)
  , lc  canFlowTo l, lc' canFlowTo l 
  = Pg lc (TReturn (TLabeled l t'))
  | otherwise 
  = Pg lc (TReturn (TLabeled l TException))

eval (Pg lc t) 
 = Pg lc (evalTerm t) 

eval PgHole    
 = PgHole  
\end{mcode}\\
&&\\
\begin{mcode}  
evalTerm :: Label l => Term l -> Term l 
evalTerm (TLabelOf (TLabeled l _)) 
  = TLabel l 
evalTerm (TLabelOf t)              
  = TLabelOf (evalTerm t)
evalTerm (TApp (TLam x t) tx)
  = subst (x,tx) t
evalTerm (TApp t tx)
  = TApp (evalTerm t) tx     
evalTerm v 
  = v     
\end{mcode} &\quad&
\begin{mcode}  
eval$*$ :: Label l => Program l -> Program l 
eval$*$ PgHole           
  = PgHole  
eval$*$ (Pg lc (TLIO t)) 
  = Pg lc (TLIO t) 
eval$*$ p                
  = eval$*$ (eval p)  
  
subst :: Eq l => (Int, Term l) 
      -> Term l -> Term l  
subst = ...
\end{mcode} \end{tabular}
\caption{Operational semantics of \liocalc.}
\label{fig:label:calculus}
\end{figure}
\Cref{fig:label:calculus} summarizes the operational semantics 
of \liocalc as three main functions, 
(i) @eval@ evaluates monadic terms taking into account the current label of 
the program,
(ii) @evalTerm@ evaluates pure terms, and 
(iii) @eval$*$@ is the transitive closure of @eval@.

\paragraph{Program evaluation}
The bind of two terms @t1@ and @t2@ fully evaluates @t1@ 
into a monadic value, 
using evaluation's transitive closure @eval$*$@. The result is 
passed to @t2@. 
The returned program uses the label of the evaluation of @t1@, 
which is safe since evaluation only increases the current label. 
In the definition of evaluation, we use Haskell's guard syntax 
@Pg lc' (TLIO t1') <- eval$*$ (Pg lc t1)@
to denote that evaluation of bind only occurs when 
@eval$*$ (Pg lc t1)@ returns a program whose term is a monadic value @TLIO@. 
Using refinement types, we prove that 
assuming that programs cannot diverge and are well-typed (\ie @t1@ is a monadic term), 
@eval$*$ (Pg lc t1)@ always returns a program with a monadic value, 
so evaluation of bind always succeeds. 
Evaluation of the @TReturn@ term simply returns a monadic value and
evaluation of @TGetLabel@ returns the current label. 
Evaluation of @TTLabel (TLabel l) t@ returns the term @t@ labeled with @l@, 
when the current label can flow to @l@, otherwise it returns an exception. 
Dually, unlabeling @TLabeled l t@ returns the term @t@ with the current 
label joined with @l@. 
The term @ToLabeled (TLabel l) t@ under current label @lc@
fully evaluates the term @t@ into a monadic value @t'@ with returned label @lc'@. 
If both the current and returned labels can flow into @l@, 
then evaluation returns the term @t@ labeled with the returned label @lc'@, 
while the current label remains the same. 
That is, evaluation of @t@ can arbitrarily raise the label, since its result 
is labeled under @l@.
Otherwise, an exception is thrown. 
The rest of the terms are pure, and their evaluation rules are given
below. Finally, evaluation of a hole is an identity.  

\paragraph{Term evaluation}
Evaluation of the term @TLabelOf t@ returns the label of @t@, if @t@ is a labeled term; 
otherwise it propagates evaluation until @t@ is evaluated to a labeled term. 
Evaluation of application uses the standard call-by-name semantics. 
The definition of substitution is standard and omitted. 
The rest of the pure terms 
are either values or a variable, whose evaluation is 
defined to be the identity. 
%
%
We define @eval$*$@ to be the transitive closure of @eval@. 
That is, @eval$*$@ repeats evaluation until a monadic value is reached. 
%

\subsection{Noninterference}\label{subsec:formal:noninterference}

Now we prove noninterference for \liocalc.
Noninterference holds when the \emph{low view}
of a program is preserved by its evaluation. This low view is 
characterized by an \emph{erasure} function, which removes program
elements whose security label is higher than the adversary's label,
replacing them with a ``hole.'' Two versions of the program 
given possibly different secrets will start with the same low view, and if the
program is noninterfering, they will end with the same low view. 
We prove nointerference of \liocalc by employing a simulation lemma,
in the style of~\citet{Li2010, RussoCH08, lio}. We use refinement
types to express this 
lemma and the property of noninterference, and rely on Liquid
Haskell to certify our proof.

\subsubsection{Erasure}
The functions @ε@ and @εTerm@ erase 
the sensitive data of programs and terms, \textit{resp.}
\begin{tabular}{lcl}
\begin{mcode}
ε     :: Label l => l -> Program l -> Program l 
εTerm :: Label l => l -> Term l    -> Term 
εTerm l (TLabeled l1 t) 
  | l1 canFlowTo l    = TLabeled l1 (εTerm l t)
  | otherwise = TLabeled l1 THole
εTerm l (TTLabel (TLabel l1) t)  
 | l1 canFlowTo l     = TTLabel (TLabel l1) (εTerm l t)
 | otherwise  = TTLabel (TLabel l1) THole 
...
\end{mcode} &&
\begin{mcode}


ε l (Pg lc t) 
 | lc canFlowTo l    = Pg lc (εTerm l t) 
 | otherwise = PgHole  
ε _ PgHole   = PgHole
$\quad$
$\quad$
$\quad$
\end{mcode}
\end{tabular}

The term erasure function @εTerm l@ replaces 
terms labeled with a label @l1@ with a hole,
if @l1@ cannot flow into the erasure label @l@.
Similarly, term erasure preemptively replaces the term @t@
in @TTLabel (TLabel l1) t@ with a hole
when @l1@ cannot flow into the erasure label @l@,
since evaluation will lead to a labeled term.
%
%
For the remaining terms, erasure is a homomorphism.
Program erasure with label @l@ of a program with current label @lc@ 
erases the term of the program, if @lc@ can flow into @l@;
otherwise it returns a program hole hiding from the attacker 
all the program configuration (\ie both the term and the current label). 
Erasure of a program hole is an identity.

\subsubsection{Simulation}
\label{subsub:simulation}

\begin{figure}
\begin{tabular}{cc}
    \begin{minipage}{.45\textwidth}
\begin{tikzpicture}
  \matrix (m) [matrix of math nodes,row sep=3em,column sep=7.5em,minimum width=2em]
  {
    \texttt{p} & \texttt{p'} \\
     \epsilon \texttt{ l p} & \epsilon \texttt{ l p'} \\};
  \path[-stealth]
    (m-1-1) edge node [left] {$\epsilon \texttt{ l}$} (m-2-1)
            edge  node [below] {$\texttt{eval}$} (m-1-2)
    (m-2-1.east|-m-2-2) edge node [below] {$\epsilon \texttt{ l . eval}$} (m-2-2)
    (m-1-2) edge node [right] {$\epsilon \texttt{ l}$} (m-2-2);
\end{tikzpicture} 
    \end{minipage}
& 
\begin{mcode}
measure terminates :: Program l -> Bool 

simulation :: Label l => l:l
  -> p:{Program l | terminates p } 
  -> { ε l (eval (ε l p)) = ε l (eval p) }
\end{mcode} 
\end{tabular}
\caption{Simulation between $\texttt{eval}$ and $\epsilon\texttt{ l . eval}$.}
\label{fig:simulation}
\end{figure}
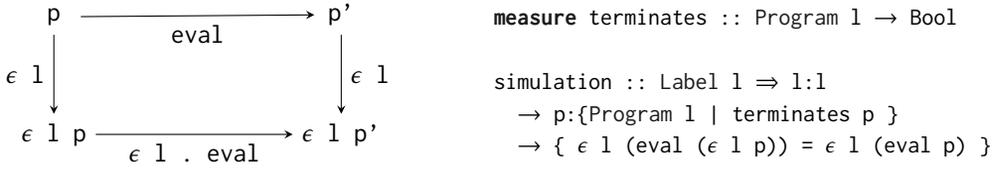
In \Cref{fig:simulation} we state that for every label @l@, 
@eval@ and @ε l . eval@ form a simulation. 
That is, evaluation of a program @p@ 
and evaluation of its erased version @ε l p@
cannot be distinguished after erasure. 
We prove this property by induction on the 
input program term. 

\paragraph{Termination}
Simulation (and later, noninterference) is termination-insensitive: it
is defined only for executions that terminate, as indicated by the
@terminates@ predicate. 
(\liocalc includes untyped lambda calculus, so \liocalc programs are
not strongly normalizing.)
This is necessary because, for soundness, Liquid Haskell 
disallows non-terminating functions, like @eval@, 
from being lifted into refinement types. 
To lift @eval@ in the logic we constrained it to only be called on terminating programs. 
To do so, we defined two logical, uninterpreted functions. 
\begin{mcode}
  measure terminates :: Program l -> Bool
  measure evalSteps  :: Program l -> Int
\end{mcode}

We use a refinement-type precondition 
to prescribe that @eval@ is only called on programs @p@ 
that satisfy the @terminates@ predicate,
and prove termination of @eval@ by checking 
that the steps of evaluation (@evalSteps p@) are decreasing at each recursive call.
\begin{mcode}
  eval :: Label l => p:{Program l | terminates p} -> Program l / [evalSteps p]
\end{mcode}
While the functions @terminates@ and @evalSteps@ cannot be defined 
as Haskell functions, we can instead \emph{axiomatize}
properties that are true under the assumption of termination. In particular,
\begin{itemize}
\item if a program terminates, so do its subprograms, and 
\item if a program terminates, its evaluation steps are strictly
  smaller than those of its subprograms. 
\end{itemize}
To express these properties, we define axioms involving these 
functions in refinements for each source program construct. 
For instance, the following assumption (encoded as a Haskell function) 
handles bind terms: 
\begin{mcode}
  assume evalStepsBindAxiom :: lc:l -> db:DB l -> t1:Term l 
   -> t2:{Term l | terminates (Pg lc db (TBind t1 t2)) } -> 
  {  (evalSteps (Pg lc db t1) < evalSteps (Pg lc db (TBind t1 t2))) 
    && (0 <= evalSteps (Pg lc db t1))
    && (terminates (Pg lc db t1))} }
  evalStepsBindAxiom _ _ _ _ = () 
\end{mcode}
Here, @evalStepsBindAxiom@ encodes that if the program 
@Pg lc db (TBind t1 t2)@ terminates, then so does @Pg lc db t1@ with fewer evaluation steps. 
This assumption is required to prove simulation in the inductive case of the @TBind@,
since we need to 
\begin{itemize}
\item apply the simulation lemma for the @Pg lc db t1@ program, 
  thus we need to know that it terminates; and 
\item prove that the induction is well founded, which we do by proving
  that the evaluation step counts of each subprogram are a decreasing
  natural number.
\end{itemize}

%

\subsubsection{Noninterference}
The noninterference theorem states that 
if two terminating \liocalc programs @p1@ and @p2@ are equal after erasure with label @l@, 
then their evaluation is also equal after erasure with label @l@. 
As with simulation, noninterference is termination
insensitive---potentially diverging programs could violate noninterference.  

We express the noninterference theorem as a refinement type.
\begin{mcode}	
  nonInterference :: Label l => l:l 
   -> p1:{Program l  | terminates p1 } -> p2:{Program l  | terminates p2 }
   -> { ε l p1 == ε l p2 } -> { ε l (eval p1) == ε l (eval p2) } 
\end{mcode}
The proof proceeds by simple rewriting using the simulation 
property at each input program and the low equivalence precondition.
\begin{mcode}    
  nonInterference l p1 p2 lowEquivalent
    =   ε l (eval p1)       ? simulation l p1  
    ==. ε l (eval (ε l p1)) ? lowEquivalent
    ==. ε l (eval (ε l p2)) ? simulation l p2  
    ==. ε l (eval p2)
    $***$ QED 
\end{mcode}
The body of @nonInterference@  
starts from the left hand side
of the equality and, using equational reasoning 
and invocation of the @lowEquivalent@ and the @simulation@ theorem 
on the input programs @p1@ and @p2@, reaches the right hand side of the equality.
As explained in~\ref{subsec:label:class} the proof combinator 
@x ? p@ returns its first argument and extends the SMT environment
with the knowledge of the theorem @p@.  
The proof combinator @x ==. y = y@ equates its two arguments and
returns the second argument to continue the equational steps. Finally, 
@x $***$ QED = ()@ casts its first argument into @unit@, 
so that the equational proof returns a @unit@ type.  

\section{Label-based Security for Database Operations}\label{sec:formal-db}
In this section we extend \liocalc with
support for databases with label-based policies. We call the extended calculus
\lwebcalc.
In~\S~\ref{subsec:database:definitions}, we define a database that stores rows 
with three values: a key, a first field with a static label, and a second field 
whose label is a function of the first field. This simplification of
the full generality of \lweb's implementation (which permits any field
to be a label) captures the key idea that fields can serve as labels
for other fields in the same row, and fields that act as labels must be
labeled as well. 
In \S~\ref{subsec:database:pure} 
we define operations to insert, select, delete, and update the database.
For each of these operations, in~\S~\ref{subsec:db:monadic}
we define a monadic term that respects the database policies. 
Finally in \S~\ref{subsec:db:noninterference}
we define erasure of the database and prove noninterference. 

\subsection{Database Definition}\label{subsec:database:definitions}
\begin{figure}[t]
\begin{tabular}{c}
\begin{mcode}  
type DB l     = [(Name, Table l)]
type Name     = String  
data Table l  = Table {tpolicy :: TPolicy l, tRows :: [Row l]}
data Row l    = Row {rKey :: Term l, rVal1 :: DBTerm l, rVal2 :: DBTerm l } 
type DBTerm l = {t:Term l | isDBValue t }                   
data TPolicy l  = TPolicy { tpTableLabel  :: l , tpFresh :: Int
                          , tpLabelField1 :: {l1:l | l1 canFlowTo tpTableLabel }
                          , tpLabelField2 :: Term l -> l }
\end{mcode}   
\end{tabular}
\caption{Definition of \lwebcalc database}
\label{fig:database-def}
\end{figure}
\Cref{fig:database-def} contains Haskell definitions used to express
the semantics of database operations in \lwebcalc. Rather than having
concrete syntax (\eg as in Figure~\ref{fig:friendstable}) for database definitions,
in our formalization we assume that databases are defined directly in
the semantic model. 

A database @DB l@ maps names (@Name@) to tables (@Table l@). 
A table consists of a policy (@TPolicy l@) and a list of
rows (@[Row l]@).
Each row contains three terms: the key and two values.
We limit values that can be stored in the database to basic terms such as
unit, integers, label values, etc. This restriction is expressed by
predicate @isDBValue@. 
Labeled terms are not permitted---labels of stored data
are specified using the table policy.
In \Cref{subsec:db:noninterference} we define erasure of the database 
to replace values with holes, thus @isDBValue@ should be true for holes too, 
but is false for any other term.

\begin{tabular}{lcl}
\begin{mcode}
isDBValue :: Term l -> Bool 
isDBValue THole    = True
isDBValue (TInt _) = True 
\end{mcode} &\quad\quad &
\begin{mcode}
isDBValue TUnit      = True 
isDBValue (TLabel _) = True
isDBValue _          = False 
\end{mcode}
\end{tabular}

We define the refinement type alias @DBTerm@ 
to be terms refined to satisfy the @isDBValue@ predicate
and define rows to contain values of type @DBTerm@. 

\paragraph{Table policy}
The table policy @TPolicy l@ defines the security policy for a table. 
The field @tpTableLabel@ is the label required to 
access the length of the table. 
The field @tpLabelField1@ is the label required to 
access the first value stored in each row of the table. 
This label is the same for each row 
and it is refined to flow into the @tpTableLabel@.
The field @tpLabelField2@ defines the  
label of the second value stored in a row as a function of the first. 
%
%
Finally, the field @tpFresh@ is used to provide 
a unique term key for each row.
The term key is an integer term that is increased at each row insertion. 

\paragraph{Helper functions}
For each field of @TPolicy@, 
we define a function that given a table 
accesses its respective policy field. 

\begin{tabular}{lcl}
\begin{mcode}
labelT   t = tpTableLabel  (tpolicy t)
labelF1  t = tpLabelField1 (tpolicy t)
\end{mcode}  & \quad &
\begin{mcode}
labelF2  t v = tpLabelField2 (tpolicy t) v
freshKey t   = tpFresh  (tpolicy t)
\end{mcode}  
\end{tabular}

We use the indexing function @db!!n@
to lookup the table named @n@ in the database.
\begin{mcode}
  (!!) :: DB l -> Name -> Maybe (Table l) 
\end{mcode}  

  
\subsection{Querying the Database}\label{subsec:database:pure}

\paragraph{Predicates}
We use predicates to query database rows. In the \lweb implementation,
predicates are written in a domain-specific query language, called
@lsql@, which the \lweb compiler can analyze. Rather than formalizing
that query language in \lwebcalc, we model predicates abstractly using
the following datatype:

\begin{mcode}
  data Pred = Pred { pVal :: Bool , pArity :: { i:Int | 0 <= i <= 2 } }
\end{mcode}  

Here, @pVal@ represents the outcome of evaluating the predicate on an
arbitrary row, and @pArity@ represents which of the row's fields were
examined during evaluation. That is, a @pArity@ value of @0@, @1@, or
@2@, denotes whether the predicate depends on (\ie computes over)
none, the first, or both fields of a row, respectively.

Then, we define a logical \textit{uninterpreted} function @evalPredicate@ that 
evaluates the predicate for some argument of type @a@:
\begin{mcode}
  measure evalPredicate :: Pred -> a -> Bool
\end{mcode}  
We define a Haskell (executable) function @evalPredicate@ and use an axiom to 
connect it with the synonymous logical uninterpreted function~\cite{refinement-reflection}:
\begin{mcode}
  assume evalPredicate :: p:Pred -> x:a -> {v:Bool | v == evalPredicate p x } 
  evalPredicate p x = pVal p    
\end{mcode}
This way, even though the Haskell function @evalPredicate p x@ returns
a constant boolean ignoring its argument @x@, the Liquid Haskell
model assumes that it behaves as an uninterpreted function that does
depend on the @x@ argument (with dependencies assumed by the @pArity@
definition). 

\paragraph{Primitive queries}
It is straightforward to define primitive operators 
that manipulate the database but do not perform IFC checks. 
We define operators to insert, delete, select, and update
databases. 
\begin{mcode}
  (+=) :: db:DB l -> n:Name -> r:Row l -> DB l                                -- insert
  (?=) :: db:DB l -> n:Name -> p:Pred  -> Term l                              -- select
  (-=) :: db:DB l -> n:Name -> p:Pred  -> DB l                                -- delete
  (:=) :: db:DB l -> n:Name -> p:Pred  -> v1:DBTerm l -> v2:DBTerm l -> DB l  -- update
\end{mcode}
\begin{itemize}[leftmargin=14.0mm]
\item[\textit{Insert:}] @db += n r@ inserts the row @r@ in the @n@ table in the database and increases @n@'s unique field. 
\item[\textit{Select:}] @db ?= n p@ selects all the rows of the @n@ table that satisfy the predicate @p@ as a list of labeled terms.
\item[\textit{Delete:}] @db -= n p@ deletes all the rows of the @n@ table that satisfy the predicate @p@.
\item[\textit{Update:}] @db := n p v1 v2@ updates each row with key @k@ of the @n@ table 
that satisfies the predicate @p@ with @Row k v1 v2@.
\end{itemize}

%
Next we extend the monadic programs of~\S~\ref{sec:formal}
with database operations to define monadic query operators 
that enforce the table and field policies.

\subsection{Monadic Database Queries}\label{subsec:db:monadic}

\begin{figure}[t]
\begin{tabular}{c}
\begin{mcode}
data Program l = 
 Pg { pLabel :: l, pDB :: DB l, pTerm :: Term l } | PgHole { pDB :: DB l } 
    
data Term l = ...
 | TInsert Name (Term l) (Term l) | TSelect Name Pred  
 | TDelete Name Pred              | TUpdate Name Pred (Term l) (Term l)
\end{mcode}  
\end{tabular}
\caption{Extension of programs and terms with a database.}
\label{fig:database:def}
\end{figure}
\subsubsection{Syntax}
\Cref{fig:database:def} defines \lwebcalc's syntax as an extension of
\liocalc. 
Programs are extended to carry the state of the database. 
Erasure of a program at an observation level @l@
leads to a @PgHole@ that now carries a database erased at level @l@.
Erasure is defined in~\S~\ref{subsec:db:noninterference};
here we note that preserving the database at program erasure
is required since even though the result of the program is erased, 
its effects on the database persist. 
For instance, when evaluating @TBind t1 t2@
the effects of @t1@ on the database affect computing @t2@. 

Terms are extended with monadic database queries. 
@TInsert n (TLabeled l1 v1) (TLabeled l2 v2)@ inserts into the table @n@ 
database values @v1@ and @v2@ labeled with @l1@ and @l2@, respectively. 
%
%
@TSelect n p@ selects the rows of the table @n@ that satisfy the predicate @p@. 
@TDelete n p@ deletes the rows of the table @n@ that satisfy the predicate @p@. 
Finally, @TUpdate n p (TLabeled l1 v1) (TLabeled l2 v2)@ updates the fields for each row 
of table @n@ 
that satisfies the predicate @p@ to be @v1@ and @v2@, 
where the database values @v1@ and @v2@ are labeled with @l1@ and @l2@, respectively. 
%

\subsubsection{Semantics}\label{subsubsec:semantics}
Figure~\ref{fig:database-eval} defines the operational semantics
for the monadic database queries in \lwebcalc.
Before we explain the evaluation rules, note that both 
insert and update attempt to insert a labeled value
@TLabeled li vi@ in the database, thus @vi@ should be a value, and
unlabeled, \ie satisfy the @isDBValue@ predicate.\footnote{We could
  allow inserting unlabeled terms, the label for which is just the
  current label. Explicit labeling is strictly more general.}
In the \lweb implementation we use Haskell's type system to enforce
this requirement. 
In \lwebcalc, we capture this property in a predicate @ς@ 
that constrains labeled values in insert and update to be database values: 

\begin{mcode}
  ς :: Program l -> Bool 
  ς (Pg _ _ t) = ςTerm t

  ςTerm :: Term l -> Bool 
  ςTerm (TInsert _ (TLabeled _ v1) (TLabeled _ v2))   = isDBValue v1 && isDBValue v2
  ςTerm (TUpdate _ _ (TLabeled _ v1) (TLabeled _ v2)) = isDBValue v1 && isDBValue v2
  ...
\end{mcode}

We specify that @eval@ is only called on \safe programs, \ie those that satisfy @ς@.
For terms other than insert and update, \safety is homomorphically
defined. Restricting \safety to permit only database values, as opposed to
terms that eventually evaluate to database values, was done to reduce
the number of cases for the proof, but does not remove any conceptual
realism. 

\begin{figure}[t]
\begin{tabular}{c}
\begin{mcode}
eval :: Label l => i:{ Program l | ς i && terminates i} -> {o:Program l | ς o } 
\end{mcode}
\end{tabular}
\begin{tabular}{ll}
\begin{mcode}
eval (Pg l db (TInsert n t1 t2) 
 | TLabeled l1 v1 <- t1
 , TLabeled l2 v2 <- t2
 , Just t <- db!!n, l1 canFlowTo labelF1 t
 , l2 canFlowTo labelF2 t v1, l canFlowTo labelT t
 = let k = freshKey t 
       r = Row k v1 v2 in 
   Pg (l join l1) (db += n r) (TReturn k)
   
eval (Pg l db (TInsert n t1 t2)) 
 | TLabeled l1 v1 <- t1
 , TLabeled l2 v2 <- t2
 = Pg (l join l1) db (TReturn TException)


eval (Pg l db (TDelete n p))   
 | Just t <- db!!n
 , l join labelPred p t canFlowTo labelT t   
 = let l' = l join labelRead p t in 
   Pg l' (db -= n p) (TReturn TUnit)
 
eval (Pg l db (TDelete n p))   
 | Just t <- db!!n  
 = let l' = l join labelRead p t in 
   Pg l' db (TReturn TException)
 | otherwise
 = Pg l  db (TReturn TException)
\end{mcode}
 &
\begin{mcode}
eval (Pg l db (TSelect n p))   
 | Just t <- db!!n 
 = let l' = l join labelT t 
              join labelPred p t in 
   Pg l' db (TReturn (db ?= n p)) 

eval (Pg l db (TSelect n p))   
 = Pg l  db (TReturn TException)   


eval (Pg l db (TUpdate n p t1 t2)   
 | TLabeled l1 v1 <- t1
 , TLabeled l2 v2 <- t2
 , Just t <- db!!n 
 , l join l1 join labelPred p t canFlowTo labelF1 t
 , l join l2 join labelPred p t canFlowTo labelF2 t v1
 = let l' = l join l1 join labelRead p t join labelT t 
   in Pg l' (db := n p v1 v2) (TReturn TUnit)
   
eval (Pg l db (TUpdate n p t1 t2)   
 | TLabeled l1 v1 <- t1
 , TLabeled l2 v2 <- t2
 , Just t <- db!!n 
 = let l' = l join l1 join labelRead p t join labelT t 
   in Pg l' db (TReturn TException)
 | otherwise 
 = Pg l  db (TReturn TException)
\end{mcode}
\end{tabular}
\caption{Evaluation of monadic database terms.}
\label{fig:database-eval}
\end{figure}

\paragraph{Insert}
Insert attempts to insert a row with 
values @v1@ and @v2@, labeled with @l1@ and @l2@ respectively,
in the table @n@. To perform the insertion we check that 
\begin{enumerate}[leftmargin=*]
\item the table named @n@ exists in the database, as table @t@. 
\item @l1@ can flow into the label of the first field of @t@, 
since the value @v1@ labeled with @l1@ will write to the first field of the table.
\item @l2@ can flow into the label of the second field of @t@, as potentially determined
by the first field @v1@ (\ie per @labelF2 t v1@).
\item the current label @l@ can flow to the label of the table, since insert changes the length of the table.
\end{enumerate}
If all these checks succeed, we compute a fresh key @k = freshKey t@,
insert the row @Row k v1 v2@ into the table @n@, and return the
key. 
If any of the checks fail we return an
exception and leave the database unchanged. 

Either way, we raise the current label @l@ by joining it with
@l1@. This is because checking @l2 $\sqsubseteq$ labelF2 t v1@
requires examining @v1@, which has label @l1@. That this check
succeeds can be discerned by whether the key is returned; if the check
fails an exception is thrown, potentially leaking information about
@v1@. This subtle point was revealed by the formalization: Our
original implementation failed to raise the current label
properly. 


\paragraph{Select}
Select only checks that the table @n@ exists in the database, 
returning an exception if it does not. 
If the table @n@ is found as the table @t@, 
then we return the term @db ?= n p@
that contains a list of all rows of @t@ that satisfy the predicate @p@, 
%
leaving the database unchanged. 
The current label is raised 
to include the label of the table @labelT t@ since on a trivially true predicate, 
all the table is returned, thus the size of the table can leak. 
We raise the current label with 
the label of the predicate @p@ on the table @t@ 
that intuitively permits reading all the values of @t@ 
that the predicate @p@ depends on. 
We define the function @labelPred p t@ that computes 
the label of the predicate @p@ on the table @t@.
\begin{mcode}
  labelPred  :: (Label l) => Pred -> Table l -> l  
  labelPred p (Table tp rs) 
   | pArity p == 2 = foldl (join) (labelF1 tp) [labelF2 tp v1 | Row _ v1 _ <- rs]
   | pArity p == 1 = labelF1 tp
   | otherwise     = bot 
\end{mcode}
If the predicate @p@ depends on both fields, 
then its predicate is the join of the label of the first field 
and all the labels of the second fields. 
If @p@ only depends on the first field, then the label of the predicate @p@ is the label of the first field. 
Otherwise, @p@ depends on no fields and its predicate is $\bot$.

Note that the primitive selection operator @db ?= n p@
returns labeled terms protected by the labels returned by the
@labelF1@ and @labelF2@ functions. 
Since terms are labeled, 
select does not need to raise the 
current label to protect values that the predicate @p@ does not read. 

\paragraph{Delete}
Deletion checks that the table named @n@ exists in the database as @t@
and that the current label joined with the label of the predicate @p@ on the table 
@t@ can flow into the label of the table @t@, since delete changes the size of the table. 
If both checks succeed, then 
database rows are properly deleted. 
The current label is raised with the ``read label'' of the predicate @p@ on the table @t@
that intuitively gives permission to read the label of the predicate @p@ on the same table. 
The function @labelRead p t@ computes the read label of the predicate @p@ 
on the table @t@ to be the label required to read @labelPredRow p t@, 
\ie equal to the label of the first field, if the predicate depends on the second field
and bottom otherwise.
\begin{mcode}
 labelRead :: (Label l) => Pred -> Table l -> l 
 labelRead p t = if pArity p == 2 then labelF1 t else bot 
\end{mcode}
Note that @labelRead p t@ always flows into @labelPred p t@, 
thus the current label is implicitly raised to this read label. 
When the runtime checks of @delete@ fail we return an exception and the database is not changed. 
If the table @n@ was found in the database, the current label is raised, even in the case of failure, 
since the label of the predicate was read. 
 
\paragraph{Update}
Updating a table @n@ with values @v1@ and @v2@ on a predicate @p@
can be seen as a select-delete-insert operation. 
But, since the length of the table is not changing, 
the check that the current label can flow to the label of the table is omitted. 
Concretely, update checks that 
\begin{enumerate}[leftmargin=*]
\item the table named @n@ exists in the database, as table @t@, 
\item @l join l1 join labelPred p t@ can flow into the label of the first field of @t@, 
since the value @v1@ labeled with @l1@ will write on the first field of the table
and whether this write is done or not depends on the label of the predicate @p@ as a hole,
\item @l join l2 join labelPred p t@ can flow into the label of the second field of @t@ when the first field is @v1@.
\end{enumerate}
If these checks succeed, then unit is returned, 
the database it updated, and the current label is raised 
to all the labels of values read during the check, \ie @l1 join labelF1 t@.
If the checks fail then we return an exception and the database is not updated.

In both cases, the current label is raised by joining with the table
label, \ie @l' = ... $\sqcup$ labelT t@. This is because the last
check depends on whether the table is 
empty or not, and its success can be discerned: if it
succeeds, then unit is returned. 
Interestingly, our original implementation failed to update the
current label in this manner. Doing so seemed intuitively unnecessary
because an update does not change the table length. 


\subsection{Noninterference}\label{subsec:db:noninterference}
As in~\S~\ref{sec:formal} to prove noninterference we prove
the simulation between @eval@ and @ε l . eval@ for \lwebcalc programs.
%
%
\begin{figure}[t]
\begin{tabular}{c}
\begin{mcode}
ε      :: (Label l) => l -> Program l -> Program l
εDB    :: (Label l) => l -> DB l -> DB l 
εTable :: (Label l) => l -> Table l -> Table l 
εRow   :: (Label l) => l -> TPolicy l -> Row l -> Row l 
\end{mcode}
\end{tabular}
\begin{tabular}{lcl}
\begin{mcode}
ε l (PgHole db) 
  = PgHole (εDB l db)
ε l (Pg lc db t) 
 | not  (lc canFlowTo l)
 = PgHole (εDB l db) 
 | otherwise 
 = Pg lc (εDB l db) (εTerm l t)

εDB l []         
 = []
εDB l ((n,t):db) 
 = (n,εTable l):εDB l db
\end{mcode}  &\quad&
\begin{mcode}
εTable l (Table tp rs) | not  (tpTableLabel tp canFlowTo l)
 = Table tp [] 
εTable l (Table tp rs)
 = Table tp (map (εRow l tp) rs) 

εRow l tp (Row k v1 v2) 
 | not  (tpLabelField1 tp canFlowTo l) 
 = Row k THole THole
 | not  (tpLabelField2 tp v1 canFlowTo l)   
 = Row k (εTerm l v1) THole 
 | otherwise                        
 = Row k (εTerm l v1) (εTerm l v2) 
\end{mcode}  
\end{tabular}
\caption{Erasure of programs and databases.}
\label{fig:database:erasure}
\end{figure}
Figure~\ref{fig:database:erasure} extends erasure to programs and databases. 
Erasure of programs is similar to~\S~\ref{sec:formal} but now we also erase the database. 
Erasure of a database recursively erases all tables. 
Erasure of a table removes all of its rows
if the label of the table cannot flow into the erasing label, thus hiding the size of the table. 
Otherwise, it recursively erases each row. 
Erasure of a row respects the dynamic labels stored in the containing
table's policy. 
Erasure of a row replaces \emph{both} fields with holes 
if the label of the first field cannot flow into the erasing label, 
since the label of the second field is not visible. 
If the label of the second field cannot flow into the erasing label, 
it replaces only the second field with a hole. 
Otherwise, it erases both fields. 

With this definition of erasure, we prove
the simulation between @eval@ and @ε l . eval@, and with this, noninterference.
The refinement properties in the database definition of \cref{fig:database-def}
are critical in the proof, as explained below.

\paragraph{\Safe programs}
The simulation proof assumes that the input program is \safe, 
\ie satisfies the predicate @ς@ as defined in~\cref{subsubsec:semantics},
or equivalently evaluation only inserts values that satisfy the @isDBValue@
property. 
To relax this assumption, an alternative approach could be to  
check this property at runtime, just before insertion of the values. 
But, this would break simulation: 
@TInsert n (TLabeled l1 v1) t@
will fail if @v1@ is not a database value, 
but its erased version can succeed if @v1@ is erased to a hole 
(when @l1@ cannot flow into the erase label).
%
%
Thus, the @isDBValue@ property cannot be checked before insertion 
and should be assumed by evaluation.
In the implementation this safety check is enforced by Haskell's 
type system. 

\paragraph{Database values}
Simulation of the delete operation requires 
that values stored in the database must have identity erasure, 
\eg cannot be labeled terms. 
Thus, we prove that all terms that satisfy @isDBValue@ 
also have erasure identity. We do this by stating the property as a 
refinement on term erasure itself.
\begin{mcode}
   εTerm :: Label l => l -> i:Term l -> {o:Term l | isDBValue i => isDBValue o }
\end{mcode}
In the delete proof, each time a database term is erased, 
the proof identity @εTerm l v == v@ is immediately available. 

\paragraph{Note on refinements}
The type @DBTerm l@ is a type alias for 
@Term l@ with the attached refinement that the term is a database value. 
%
A @DBTerm l@ \textit{does not carry} an actual proof that it is a database value. 
Instead, the refinement type that the term satisfies the @isDBValue@ property is statically verified during type checking. 
%
As a consequence, comparison of two 
@DBTerm@s does not require proof comparison.
At the same time, verification can use the @isDBValue@ property.
For instance, when opening a row @Row k v1 v2@, 
we know that @isDBValue v1@ and by the type of term erasure, 
we know that for each label @l@, @εTerm l v1 == v1@.


\section{Liquid Haskell for Metatheory}\label{sec:liquidhaskell-discussion}

Liquid Haskell was originally developed
to support lightweight program verification (\eg
out-of-bounds indexing). 
The formalization of \lweb in Liquid Haskell, presented
in~\Cref{sec:formal} and~\Cref{sec:formal-db}, was made possible by
recent extensions to support general theorem
proving~\cite{refinement-reflection}. Our proof of 
noninterference was a challenging test of this new support, and
constitutes the first advanced metatheoretical result mechanized in
Liquid Haskell.\footnote{\url{https://github.com/plum-umd/lmonad-meta}}

The trusted computing base (TCB) of any Liquid Haskell proof relies on
the correct implementation of several parts. In particular, we trust that 
\begin{enumerate}
\item the GHC compiler correctly desugars the Haskell code to the core language 
of Liquid Haskell, 
\item Liquid Haskell correctly generates the verification conditions for the core language, and 
\item the SMT solver correctly discharges the verification conditions.  
\end{enumerate}
We worked on the noninterference proof, on and off, for 10 months. 
The proof consists of 5,447 lines of code and requires about 5 hours to be checked. 
For this proof in particular, we (naturally) trust all of our semantic
definitions, and also two explicit assumptions, notably the axiomatization of
termination and modeling of predicates. These were discussed respectively in
\cref{subsub:simulation} and \cref{subsec:database:pure}.

Carrying out the proof had a clear benefit: As mentioned in
\cref{subsec:db:monadic}, we uncovered two bugs in our
implementation. In both cases, \lweb was examining sensitive data when
carrying out a security check, but failed to raise the current label
with the label of that data. Failure of the mechanized proof to go
through exposed these bugs.

The rest of this section summarizes what we view as the (current)
advantages and disadvantages of using Liquid Haskell as a theorem
prover compared to other alternatives (e.g., Coq and
F-star~\citep{fstar}), expanding on a prior assessment~\cite{a-tale}.

\subsection{Advantages}

As a theorem proving environment, Liquid Haskell offers several
advantages. 

\paragraph{General purpose programming language.}
The Liquid Haskell-based formal development is, in essence, a
Haskell program. All formal definitions (presented in
\Cref{sec:formal} and~\Cref{sec:formal-db})
and proof terms (\eg illustrated in
\cref{subsec:formal:noninterference}) are Haskell code. 
Refinement types define lemmas and theorems, referring to these
definitions. 
In fact, some formal definitions 
(\eg the @Label@ class definition) were taken 
directly from the implementation. 
The first author of the paper and main developer of the proof is a Haskell programmer, 
thus he did not need to learn a new programming language (\eg Coq) to develop the formal proof. 
During development we used Haskell's existing development
tools, including the build system, test frameworks, and deployment
support (\eg Travis integration).

\paragraph{SMT automation}
Liquid Haskell, like Dafny~\citep{Leino:2010} and F-star~\citep{fstar}, 
uses an SMT solver to automate parts of the proof, 
especially the ones that make use of boolean reasoning, reducing the
need for manual case splitting.
For example, proving simulation for row updates normally proceeds by
case splitting on the relative can-flow-to relation between  
four labels. The SMT automates the case splitting. 

\paragraph{Semantic termination checking}
To prove termination of a recursive function in Liquid Haskell it suffices to 
declare a non negative integer value that is decreasing at each recursive call. 
%
The \lweb proof was
greatly simplified by the semantic termination checker.  
In a previous Coq LIO proof~\cite{stefan:2017:flexible}, 
the evaluation relation apparently requires an explicit \emph{fuel}
argument to count the number of evaluation steps, since the evaluation
function (the equivalent to that in \cref{fig:label:calculus}) does
not necessarily terminate. 
In our proof, termination of evaluation was axiomatized (per~\Cref{subsub:simulation}), 
which in practice meant
that the evaluation steps were counted only in the logic and not in the definition of the 
evaluation function.

\paragraph{Intrinsic \emph{and} extrinsic verification}
The Liquid Haskell proving style allows us to conveniently switch between 
(manual) extrinsic and (SMT automated) intrinsic
verification. 
Most of the \lweb proof is extrinsic, 
\ie functions are defined to state and prove theorems about the model. 
In few cases, intrinsic specifications are used to ease the proof. 
For instance, the refinement type specification of 
@εTerm@, as described in~\ref{subsec:db:noninterference},
intrinsically specifies that erasure of @isDBValue@ terms returns 
terms that also satisfy the @isDBValue@ predicate. 
This property is automatically proven by the SMT without cluttering
the executable portion of the definition with proof terms.  

\subsection{Disadvantages}

On the other hand, Liquid Haskell has room to improve as a
theorem proving environment, especially compared to 
advanced theorem provers like Coq.

\paragraph{Unpredictable verification time}
The first and main disadvantage is the unpredictability of
verification times, which owe to the invocation of an SMT solver. One
issue we ran across during the development of our proof is that
internal transformations performed by @ghc@ can cause massive blowups.
This is because Liquid Haskell analyzes Haskell's intermediate code
(@CoreSyn@), not the original source. As an example of the problem, using @|x,y@
instead of the logical @| x && y@ in function guards leads to much slower
verification times. While the two alternatives have exactly the same semantics, 
the first case leads to exponential expansion of the intermediate
code. 

\paragraph{Lack of tactics}
Liquid Haskell currently provides no tactic support, 
which could simplify proof scripts. For example, we often had to systematically
invoke label laws (\cref{fig:formalism:label}) in our proofs, whereas
a proof tactic to do so automatically could greatly simplify these cases. 

\paragraph{General purpose programming language}
Liquid Haskell, developed for light-weight verification of Haskell programs, 
lacks various features in verification-specific systems, such as Coq. 
For example, Liquid Haskell provides only
experimental support for curried, higher-order functions,  
which means that one has to inline higher order functions, like @map@, @fold@, and @lookup@.
There is also no interactive proof environment or (substantial) proof libraries.

\bigskip

In sum, our \lweb proof shows that Liquid Haskell can be used for
sophisticated theorem proving. We are optimistic that current
disadvantages can be addressed in future work.

\section{Implementation}\label{sec:impl}

\lweb has been available online since 2016 and consists of 2,664 lines
of Haskell code.\footnote{\url{https://github.com/jprider63/lmonad-yesod}}
It depends on our base \lmonad package that implements the @LMonadT@
monad transformer and consists of 345 lines of code.\footnote{\url{https://github.com/jprider63/lmonad}}
\lweb also imports 
\yesod, a well established, external Haskell library for type-safe,
web applications. 
This section explains how the implementation extends the
formalization, and then discusses the trusted computing base.

\subsection{Extensions}\label{subsec:impl:ext}

The \lweb implementation generalizes the formalization of
Sections~\ref{sec:formal} and~\ref{sec:formal-db} in several ways. 

\paragraph*{Clearance label}
The implementation supports a \emph{clearance} label, 
described in \cref{sec:lio-intro}. Intuitively, the clearance label
limits how high the current label can be raised.  If the current label
ever exceeds the clearance label, an exception is thrown.  This label
is not needed to enforce noninterference, but serves as an
optimization, cutting off transactions whose current label rises to
the point that they are doomed to fail. Adding checks to handle the
clearance was straightforward.

\paragraph*{Full tables and expressive queries} As first illustrated in \Cref{subsec:lweb}, tables may have
more than two columns, and a column's label can be determined by other
various fields in the same row. The labels of such
\emph{dependency fields} must be constant, \ie not determined by
another field, and flow into the table label (which also must be
constant). 
A consequence of this rule is that a field's label cannot depend on itself. 
Finally,
values stored in tables instantiate \yesod's @PersistField@ type
class.  The implementation uses only the predefined instances
including @Text@, @Bool@, @Int@ but critically, does not define a
@PersistField@ for labeled values. \lweb enforces these invariants at
compile time via Haskell type checking and when preprocessing table
definitions. \lweb rewrites queries to add
labels to queried results. 



We have implemented database operations 
beyond those given in \cref{sec:formal-db}, to be more in line with
typical database support. 
Some of these operations are simple variations of the ones presented.  For example, \lweb allows
for variations of @update@ that only update specific fields (not whole
rows).  \lweb
implements these basic queries by wrapping Persistent~\cite{yesod},
\yesod's database library, with the derived IFC 
checks.  To support
more advanced queries, \lweb defines an SQL-like domain-specific language called
@lsql@.  @lsql@ allows users to write expressive SQL queries that
include inner joins, outer joins, @where@ clauses, orderings, limits,
and offsets.  Haskell expressions can be included in queries using
anti-quotation.  At compile-time, \lweb parses @lsql@ queries using
quasi-quotation and Template Haskell \cite{Sheard:2002:TMH:636517.636528}.  It rewrites the queries to be
run using Esqueleto~\cite{esqueleto}, a Haskell library that supports
advanced database queries. As part of this rewriting, \lweb inserts IFC checks for
queries based on the user-defined database policies. We show several
examples of @lsql@ queries in \cref{subsec:bififi}.

\paragraph*{Optimizations}
Sometimes a label against which to perform a check is
derived from data stored in every row. Retrieving every
row is especially costly when 
the query itself would retrieve only a fraction of them. Therefore,
when possible we compute an upper bound for such
a label. In particular, if a field is fully constrained by a 
query's predicate, we use the field's constrained value to compute any
dependent labels. When a field is not fully constrained, we conservatively set
dependent labels to @top@. Suppose we wish to query the @Friends@
table from \cref{fig:friendstable}, retrieving all rows such that 
@user1 == 'Alice'@ and @date < '2000-01-01'@. The confidentiality
portion of @user1@'s label is
@bot@, but that
portion of @date@'s is computed from @user1 meet user2@. 
Since @user1@ is always @'Alice'@ we know the computed label is
$\bigsqcup_l$ @Alice@ $\sqcap ~ l$ for all values @user2 = @$l$ in the
database. In this case, we can bound $l$ as @top@, and thus use label
@Alice@, since it is equivalent to @Alice meet top@. While this bound
is technically conservative, in practice we find it makes
policy sense. In this example, if the @user2@ field can truly vary
arbitrarily then $\bigsqcup_l ~ l$ will approach @top@. 

\paragraph*{Declassification}
\lweb supports forms of 
\emph{declassification} \cite{Sabelfeld:2009:DDP:1662658.1662659} for cases when the IFC lattice ordering
needs to be selectively relaxed. These should be used sparingly (and
are, in our \bibifi case study), as they form part of the trusted
computing base, discussed below.

\paragraph*{Row ordering}
As a final point, we note that our formalization models a database as
a list of rows; insertion (via @+=@) simply appends to the
list, regardless of the contents of a row. As such, \emph{row
  ordering} does not depend on the database's contents and thus reveals
nothing about them (it is governed only by the table label).
In the implementation, advanced operations may specify an ordering. 
\lweb prevents leaks in this situation by raising the current label with the label of fields used for sorting. 
If a query does not specify an ordering, \lweb takes no specific
steps. However, ordering on rows is undefined in SQL, so
a backend database could choose to order them by their contents, and
thus potentially leak information in a query's results. 
In our experience with PostgreSQL, default row ordering 
depends on when values are written 
and 
is independent of the data in the table. 


\subsection{Trusted Computing Base}\label{subsec:impl:limitations}

A key advantage of \lweb is that by largely shifting security checks
from the application into the \lweb IFC framework, we can shrink an
application's trusted computing base (TCB). In particular, for an
application that uses \lweb, the locus of trust is on \lweb itself,
which is made (more) trustworthy by our mechanized noninterference
proof. A few parts of the application must be trusted, nevertheless.

First, all of the policy specifications are trusted. The policy
includes the labels on the various tables and the labels on data
read/written from I/O channels. Specifying the latter requires 
writing some trusted code to interpret data going in or out. For
example, in a multi-user application like \bibifi, code performing
authentication on a particular channel must be trusted
(\cref{subsubsec:users}).

Second, any uses of declassification are trusted, as they constitute
local modifications to policy. 
One kind of declassification can occur selectively
on in-application data~\cite{sabelfeld:survey}. We give an example in
\cref{subsec:bibifi:declassification}.  
Another kind of declassification is to relax some security checks during database updates. 
The update query imposes strong runtime checks, \eg
that the label of the predicate should flow into the updated fields as
formalized in~\cref{sec:formal-db}. 
LWeb
provides an unsound update alternative (called
@updateDeclassifyTCB@)
that ignores this specific check. 

%
%
%
%
%
%
%
%
%


\section{The \bibifi Case Study}\label{subsec:bififi}
As a real world web application of \lweb, we present \emph{Build it,
  Break it, Fix it} (\bibifi), a security-oriented programming
contest~\cite{Ruef:2016:BBF:2976749.2978382} hosted at
\url{https://builditbreakit.org}.  The contest consists of three
rounds.  At the outset, the organizers publish the specification for
some software that has particular security goals. During the first round,
teams implement software to this specification, aiming
for it to be both fast and secure. In the second round, teams find 
as many breaks as possible in the implementations submitted by
other teams.  During the final round, teams attempt to fix the
identified problems in their submissions.

\subsection{\bibifi Labels}
\begin{wrapfigure}{r}{0.38\textwidth}
\vspace*{-.4in}
\begin{mcode}
data Principal
 = PSys | PAdmin | PUser UserId 
 | PTeam  TeamId | PJudge JudgeId

type BBFLabel = DCLabel Principal
\end{mcode}
\caption{\bibifi labels.}
\label{code:BBFLabel}
\end{wrapfigure}
\bibifi labels include all entities that operate in the system. 
The @Principal@ data type, defined in ~\cref{code:BBFLabel},
encodes all such entities, including
the system itself,
the administrator, users, teams, and judges.
Each of these entities is treated as a security level. 
For instance a policy can encode that 
data written by a user with id @5@, 
can get protected at the security level of this specific user, 
so that only he or she can read this data. 
A more flexible policy encodes that the system administrator can read 
data written by each user. 
To encode such policies, 
we use disjunction category labels (@DCLabel@)~\cite{stefan:dclabels} to 
create a security lattice out of our @Principal@s.
In~\cref{code:BBFLabel} we define @BBFLabel@
as the @DCLabel Principal@ data type that tracks the security level of values 
as they flow throughout the web application and database.

%

\begin{wrapfigure}{r}{0.45\textwidth}
\vspace*{-.35in}
\begin{mcode}
*User*
  *account Text*
  *email Text*   ^<Const Admin meet Id, Id>^
  *admin Bool*   ^<bot, Const Admin>^
\end{mcode}
\vspace*{-.05in}
\caption{Basic \bibifi \texttt{User} table.}
\label{code:usertable}
\vspace*{-.1in}
\end{wrapfigure}

\begin{figure}[t]
\centering
\begin{minipage}{.75\textwidth}
\begin{mcode}
*UserInfo*  
  *user       UserId*
  *school     Text*  ^<Const Admin meet Field user, Field user>^
  *age        Int*   ^<Const Admin meet Field user, Field user>^
  *experience Int*   ^<Const Admin meet Field user, Field user>^
\end{mcode}
\end{minipage}
\caption{Table \texttt{UserInfo} contains additional \bibifi user information.}
\label{code:userinfotable}
\end{figure}

\subsection{Users and Authentication}\label{subsubsec:users}
Users' personal information is stored in the \bibifi database. 
\Cref{code:userinfotable}
shows the @User@ table with the basics: a user's account id,
email address, and 
whether they have administrator privileges. The label for the @email@
field refers to @Id@ in its label: This is a shorthand for the key of
the present table. The label says that a user can read and write their
emails, while the administrator can read every user's email.  The
label for the @admin@ field declares that it may be written by the
administrator and read by anyone.

Additional private information is stored in the @UserInfo@ table,
shown in \Cref{code:userinfotable}, 
including a user's school, age, and professional experience. The
@user@ field of this table is a foreign key to the @User@ table, as
indicated by its type @UserId@ (see \cref{sec:yesod}). Each of the
remaining fields is protected by this field, in part:
users can read and write their own
information while administrators can read any users' information.

The current label is set by the code trusted to perform authentication. 
If a user is not logged in, the current label is set to @<bot,top>@:
the confidentiality label is the upper bound on data read so far
(i.e., none, so @bot@), and the integrity label is the level of least
trust (i.e., @top@) for writing data.
After authenticating, most users will have the label @<bot, PUser userId>@, 
thus lowering the integrity part (thus increasing the level
of trust) to the user itself. 
Users who are also administrators will have current label lowered
further to @<bot, PUser userId meet PAdmin>@. 
This is shown in the following code snippet. It determines the logged
in user via @requireAuth@, and then adds administrator privileges if
the user has them (per @userAdmin@).
\begin{mcode}
  (Entity userId user) <- requireAuth
  let userLabel = dcIntegritySingleton (PrincipalUser userId)
  lowerLabelTCB $\$$ if userAdmin user
                     then userLabel meet dcIntegritySingleton PrincipalAdmin
                     else userLabel
\end{mcode}
The clearance is also set using trusted functions during
authentication. For example, for an adminstrator it would be 
@<PUser userId join PAdmin,top>@.

\subsection{Opening the Contest}
\begin{figure}[t]
\begin{tabular}{lclcl}
\begin{mcode}
*Announcement* ^<bot, Const Admin>^
  *title   Text* ^<bot Const Admin>^
  *content Text* ^<bot, Const Admin>^
\end{mcode} &\quad \quad\quad&
\begin{mcode}
*Team*
  *name    Text*
  *contest ContestId*
\end{mcode}
&\quad\quad\quad&
\begin{mcode}
*TeamMember*
  *team TeamId*
  *user UserId*
\end{mcode}
\end{tabular}
\caption{Definition of \texttt{Announcement}, \texttt{Team}, and
  \texttt{TeamMember} tables and their policies.}
\label{code:teamtable}
\label{code:announcementtable}
\end{figure}
To start a contest, administrators write announcements that include information like instructions and problem specifications. It is important that only administrators can post these announcements. 
Announcements are stored in the database, and their (simplified) table definition is shown in \cref{code:announcementtable}. 
The @Announcement@ table has two @Text@ fields corresponding to an announcement's title and content.
Only administrators can author announcements.

An earlier version \bibifi relied on manual access control checks 
rather than monadic \lmonad enforcement of security. 
The old version had a security bug: it failed to check that the current user was
an administrator when posting a new announcement.
Here is a snippet of the old code.
\begin{code}
  postAddAnnouncementR :: Handler Html
  postAddAnnouncementR = do
      ((res, widget), enctype) <- runFormPost postForm
      case res of ...
          FormSuccess (FormData title markdown) -> do
              runDB (insert (Announcement title markdown))
              redirect AdminAnnouncementsR
\end{code}
This function parses POST data and inserts a new announcement. 
The user is never authenticated, so anyone can post new announcements and potentially deface the website.
In the IFC version of the website, the database insertion fails for
unauthorized or unauthenticated users as the integrity part of the
current label is not sufficiently trusted (the label does not flow
into @PAdmin@).

\subsection{Teams and Declassification}\label{subsec:bibifi:declassification}
To participate in a contest, a user must join a team. 
The teams and their members are stored in the eponymous tables of~\cref{code:teamtable}. 
Teams serve as another principal in the \bibifi system 
and \bibifi defines a TCB function that appropriately 
authenticates team members similarly to users
(\cref{subsubsec:users}), authorizing a team member to read and write
data labeled with their team.

\bibifi uses declassification (as discussed in~\ref{subsec:impl:limitations}) 
to allow team members to send email messages to their team. 
The policy on the email field of the @User@ table states that only the user or an administrator can read the email address, so \bibifi cannot give a user's email address to a teammate. 
Instead, the function @sendEmailToTeam@ below 
sends the email on the teammate's behalf using declassification. 
\begin{code}
  sendEmailToTeam :: TeamId -> Email -> LHandler ()
  sendEmailToTeam tId email = do
    protectedEmails <- runDB [lsql| pselect User.email from User inner jjoin TeamMember on TeamMember.user == User.id where TeamMember.team == #{tId} |]
    mapM_ (\protectedEmail -> do
        address <- declassifyTCB protectedEmail
        sendEmail address email
      ) protectedEmails
\end{code}
The function @sendEmailToTeam@'s parameters are the team identifier
and an email return address.
It queries the database for the (labeled) email addresses of the team's
members, using @lsql@
(see~\cref{subsec:lweb} and~\cref{subsec:impl:ext}). 
The @sendEmailToTeam@ function maps over each 
address, declassifying it via @declassifyTCB@, so that the message can
be sent to the address. 
The @declassifyTCB@ function takes a labeled value and extracts
its raw value, \emph{ignoring label restrictions}. 
This is an unsafe operation that breaks noninterference, 
so the programmer must be careful with its use. 
Here for example, the function
is careful not to reveal the email address to the sender but only use
it to send the email. 
%
%

\subsection{Breaks and Advanced Queries}
During the second round of the \bibifi context, 
teams submit breaks, \ie test cases that attack another team's submission. 
After a break is pushed to a registered git repository,
\bibifi's backend infrastructure uploads 
it to a virtual machine and tests whether the attack succeeds. 
Results are stored in the @BreakSubmission@ table of~\cref{code:breaktable}, 
which has fields for the attacking team, 
the target team, and the (boolean) result of the attack. 
The integrity label for the result field is @PSys@ since only the backend system can grade an attack. 
The confidentiality label is @PAdmin meet PTeam attackerId meet PTeam targetId@ 
since administrators, the attacker team, and the target team can see the result of an attack.
\begin{figure}
\begin{mcode}
  *BreakSubmission*
    *attacker TeamId* ^<bot, Const Sys>^
    *target   TeamId* ^<bot, Const Sys>^
    *result   Bool*   ^<Const Admin meet Field attacker meet Field target, Const Sys>^
\end{mcode}
\caption{Definition of \texttt{BreakSubmission} table and its policy.} 
\label{code:breaktable}
\end{figure}

\bibifi has an administration page that lists all break submissions next to which team was attacked. 
This page's contents are retrieved via the following inner join. 
\begin{mcode}
  runDB $\$$ [lsql| select BreakSubmission.$\star$, Team.name from BreakSubmission inner jjoin Team on BreakSubmission.target == Team.id where Team.contest == #{contestId} order by BreakSubmission.id desc |]
\end{mcode}
\JPH{Could add attacking team too}
This query performs a join over the @BreakSubmission@ and @Team@ tables, aligning rows where the target team equals the team's identifier. 
In addition, it filters rows to the specified contest identifier
and orders results by the break submission identifiers. 
\mwhh{Queries: All breaks on me, all breaks I did that were successful. All
breaks by team 1 on team 2 (admin).}

\section{Experimental Evaluation}\label{sec:experiments}
To evaluate \lweb we compare the \bibifi implementation that 
uses \lmonad with our initial \bibifi implementation 
that manually checked security policies via access control. 
We call this initial version the 
\textit{vanilla implementation}.
Transitioning from the vanilla to the \lweb implementation 
reduced the trusted computing base (TCB) but imposed a modest runtime overhead. 

\subsection{Trusted Computing Base of \bibifi}
The implementation of the \bibifi application is currently 
11,529 lines of Haskell code. 
80 of these lines 
invoke trusted functions (for authentication or declassification, see~\cref{subsec:impl:limitations}).
\lweb's library is 3,009 lines of trusted code. 
The vanilla implementation is several years old, with 7,367 LOC;
there is no IFC mechanism so the whole 
codebase is trusted.
Switching from the vanilla to the \lweb implementation only added 151
LOC. The size of the TCB is now \tcbnumber of the codebase; 
considering only the code of the \bibifi web application (and not \lweb too), \tcbnumberbibifi of the code is trusted. 

\subsection{Running Time Overhead}
\begin{table}
\caption{
Latency comparison between the \textbf{Vanilla} 
and \textbf{\lweb} implementations
of the \bibifi application.  
\EDIT{3}{The mean, standard deviation, and tail latency in milliseconds over 1,000 trials are presented. In addition, the response size in kilobytes and the overhead of LWeb are shown.}
%
}
\begin{center}
\resizebox{\linewidth}{!}{
\begin{tabular}{| l | c | r | r | r | r | r | r | c | c |}
\hline
\textbf{Handler} & \textbf{Verb} & \multicolumn{3}{|c|}{\textbf{Vanilla Latency}} & \multicolumn{3}{|c|}{\textbf{LWeb Latency}} & \textbf{Size (kB)} & \textbf{Overhead} \\
\hline
				& & \textbf{Mean (ms)} & \textbf{SD (ms)} & \textbf{Tail (ms)} & \textbf{Mean (ms)} & \textbf{SD (ms)} & \textbf{Tail (ms)} & & \\
\cline{3-8}
/announcements & GET & 4.646 & 1.215 & 16 & 5.529 & 1.367 & 20 & 18.639 & 19.01\% \\
/announcement/update & POST & 9.810 & 2.600 & 54 & 11.395 & 3.054 & 52 & 0.706 & 16.16\% \\
/profile & GET & 2.116 & 0.512 & 6 & 2.167 & 0.550 & 6 & 7.595 & 2.41\% \\
/buildsubmissions & GET & 6.364 & 1.251 & 17 & 7.441 & 1.706 & 22 & 14.434 & 16.92\% \\
/buildsubmission & GET & 28.633 & 2.772 & 52 & 30.570 & 3.477 & 75 & 9.231 & 6.76\% \\
/breaksubmissions & GET & 41.758 & 7.826 & 81 & 49.218 & 11.679 & 90 & 60.044 & 17.86\% \\
/breaksubmission & GET & 4.070 & 0.538 & 9 & 4.923 & 0.509 & 9 & 6.116 & 20.96\% \\
%
\hline
\end{tabular}
}
\end{center}
\label{table:time:bibifi}
\end{table}
%
%
We measured the query latency, 
\ie the response time (in milliseconds) of HTTP requests, 
for both the \lweb and the vanilla implementation. 
\EDIT{3}{
Measurements were performed over} @localhost@ \EDITCOLOR{and we ran 100 requests to warm up. 
We present the mean, standard deviation, and tail latency over 1,000 trials, as well as the response size (in kilobytes)
and the overhead of 
\lweb over the vanilla implementation. 
\Cref{table:time:bibifi} summarizes this comparison.
The server used for benchmarking runs Ubuntu 16.04 with two Intel(R) Xeon(R) E5-2630 2.60GHz CPUs and 64GB of RAM. 
PostgreSQL 9.5.13 is run locally as the database backend.
We used ApacheBench to perform the measurements with a concurrency level of one. 
Here is a sample invocation of} @ab@:
\begin{mcode}
ab -g profile_lweb.gp -n 1000 -T "application/x-www-form-urlencoded; charset=UTF-8" -c 1 -C _SESSION=... http://127.0.0.1:4000/profile
\end{mcode}
Most of the requests are GET requests 
that display contest announcements, retrieve a user's profile with personal information, get the list of a team's submissions, and view the results of a specific submission. 
One POST request is measured that updates the contents of an announcement. 
Cookies and CSRF tokens were explicitly defined so that a user was logged into the site, and the user had sufficient permissions for all of the pages.

\begin{figure}
\centering
\begin{subfigure}{0.475\textwidth}
\includegraphics[width=\textwidth]{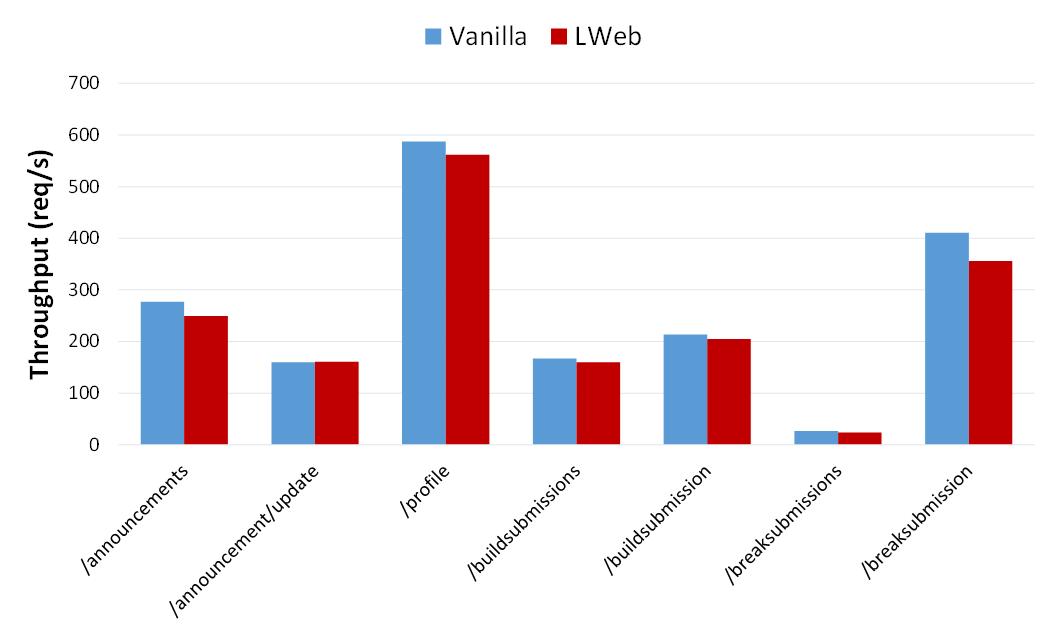}
\caption{Concurrency level of 16.}
\end{subfigure}
\hfill
\begin{subfigure}{0.475\textwidth}
\includegraphics[width=\textwidth]{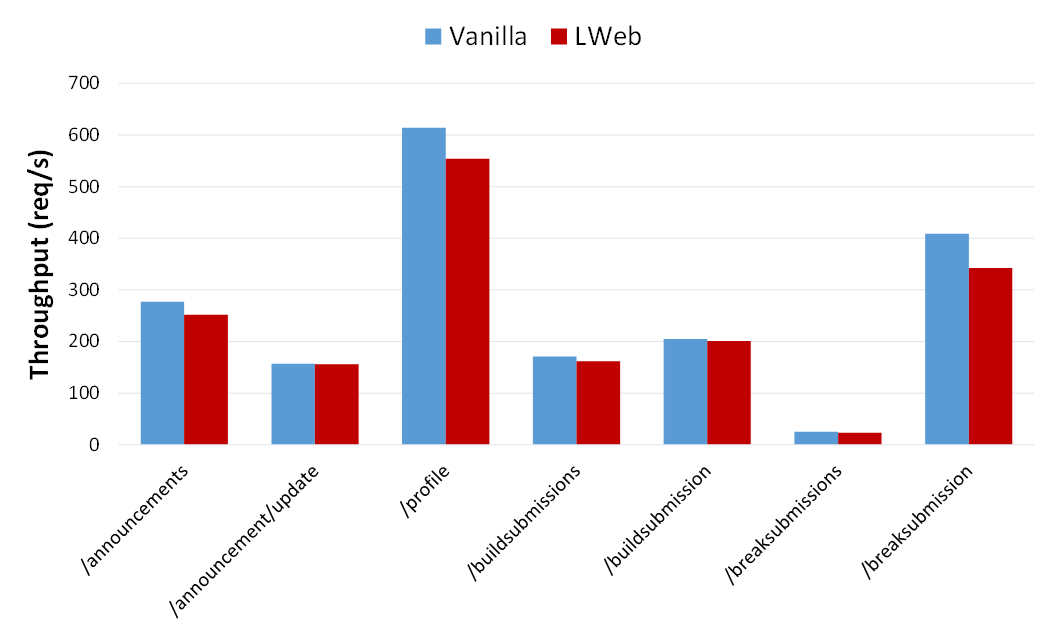}
\caption{Concurrency level of 32.}
\end{subfigure}
\caption{\EDIT{3}{Throughput (req/s) of the \textbf{Vanilla} and \textbf{\lweb} versions of the \bibifi application.}} 
\label{fig:throughput:bibifi}
\end{figure}

\EDIT{3}{To evaluate \lweb's impact on the throughput of web applications, we conduct similar measurements except we rerun} @ab@ \EDITCOLOR{with concurrency levels of 16 and 32. 
The rest of the experimental setup matches that of the latency benchmark, including number of requests, hardware, and handlers. 
\Cref{fig:throughput:bibifi} 
shows the number of requests per second for each version of the \bibifi web application
across the various handlers.
}

\EDIT{3}{Most of the handlers show modest overhead between the vanilla and \lweb versions of the website.
We measure \lweb's overhead to range from \overheadnumbermin to \overheadnumber,
which comes from 
the IFC checks that \lweb makes for every database query
and the state monad transformer that tracks the current label and clearance label. 
In practice, this overhead results in a few milliseconds of delay in response times. 
In most situations, this is a reasonable price to pay in order the reduce the size of the TCB
and increase confidence that the web application properly enforces the user defined security policies.
}

\section{Related Work}\label{sec:related}

\lweb provides end-to-end information flow control (IFC) security for
webapps. Its design aims to provide highly expressive policies and
queries in a way that does not compromise security, and adds little
overhead to transaction processing, in both space and time. This
section compares \lweb to prior work, arguing that it occupies a
unique, and favorable, spot in the design space.

\paragraph*{Information flow control}
\lweb is part of a long line of work on using lattice-ordered,
label-based IFC to enforce security policies in
software~\cite{lapadula1973,denning,sabelfeld:survey}. Enforcement can
occur either \emph{statically} at compile-time, \eg as part of type
checking~\cite{jif,flowcaml,Lourenco:2015:DIF:2676726.2676994,Lourenco:2013:IFA:3092395.3092410}
or a static
analysis~\cite{Hammer:2009:FCO:1667545.1667547,JohnsonWMC2015,Arzt:2014:FPC:2666356.2594299},
or \emph{dynamically} at run-time, \eg via source-to-source
rewriting~\cite{Chudnov:2015:IIF:2810103.2813684,hedin15jsflow} or
library/run-time
support~\cite{Roy:2009:LPF:1542476.1542484,Tromer:2016:DII:2897845.2897888,lio}. Dynamic
approaches often work by 
rewriting a program to insert the needed checks and/or by relying on
support from the hardware, operating system, or run-time.  Closely
related to IFC, \emph{taint tracking} controls \emph{data flows}
through the program, rather than overall influence (which includes
effects on \emph{control flow}, \ie \emph{implicit} flows). Taint
tracking can avoid the false positives of IFC, which often
overapproximates control channels, but will also miss security
violations~\cite{king08implicit}.

\lweb builds on the \lio framework~\cite{lio}, which is a dynamic
approach to enforcing IFC that takes advantage of Haskell's static
types to help localize checks to I/O boundaries. \lio's \emph{current
  label} and \emph{clearance label} draw inspiration from work on
Mandatory Access Control (MAC) operating systems~\cite{lapadula1973}, 
including Asbestos~\cite{Efstathopoulos:2005:LEP:1095810.1095813},
HiStar~\cite{Zeldovich:2006:MIF:1267308.1267327}, and Flume~\cite{Krohn:2007:IFC:1294261.1294293}.
The baseline \lio approach has been extended in several interesting
ways~\cite{Russo:2015:FPT:2784731.2784756,Buiras:2015:HMS:2784731.2784758, Waye:2017:CSI:3133956.3134036, Buiras13},
including to other languages~\cite{Heule:2015:IIR:2976888.2976892}. 



The proof of security in the original \lio (without use of a database)
has been partially mechanized in Coq~\cite{stefan:2017:flexible}, while the derivative MAC
library~\cite{Russo:2015:FPT:2784731.2784756} has been mechanized in
Agda~\cite{Vassena:2016:FIC:2993600.2993608}. The MAC mechanization
considers concurrency, which ours does not. Ours is the first mechanization to use an
SMT-based verifier (Liquid Haskell).

\paragraph*{IFC for database-using web applications}
Several prior works apply IFC to web applications.
FlowWatcher~\cite{Muthukumaran:2015:FDA:2810103.2813639} enforces
information flow policies within a web proxy, which provides the
benefit that applications need not be retrofitted, but limits the
granularity of policies it can enforce.

SeLINQ~\cite{Schoepe:2014:STI:2628136.2628151} is a static IFC system
for F\# programs that access a database via language-integrated
queries (LINQ). 
SIF~\cite{Chong:2007:SEC:1362903.1362904} uses Jif~\cite{jif} to
enforce static IFC-based protection for web servlets, while
Swift~\cite{Chong:2007:SWA:1294261.1294265} 
also allows client-side (Javascript) code.
Unlike \lweb, these systems permit only statically determined database
policies, not ones with dynamic labels (\eg stored in the database).
The latter two lack language support for
database manipulation, though a back-end database can be made
accessible by wrapping it with a Jif signature (which we imagine would
require an SeLINQ-style static policy).

UrFlow~\cite{urflow} performs static analysis to prove that
information flow policies are properly enforced. These policies are
expressed as SQL queries over protected data and known
information. Static analysis-based proofs about queries and flows impose
no run-time overhead. But static analysis can be overapproximate, rejecting correct
programs. Dynamic enforcement schemes do not have this issue, and
\lweb's \lio-based approach imposes little run-time overhead.

SELinks~\cite{corcoran09selinks} enforces security policies for web
applications, including ones resembling the field-dependent policies
we have in \lweb. To improve performance, security policy checks
were offloaded to the database as stored procedures; \lweb could
benefit from a similar optimization. SELinks was originally based on a formalism
called Fable~\cite{swamy08fable} in which one could encode IFC
policies, but this encoding was too onerous for practical use, and not
present in SELinks, which was limited to access control
policies. Qapla~\cite{qapla} also supports rich policies, 
but like SELinks these focus on access control, and so may fail to plug leaks
of protected data via other server state.

Jacqueline~\cite{YangHASFC16} uses faceted information
flow control~\cite{Austin:2012:MFD:2103656.2103677} to implement
policy-agnostic security~\cite{YangYS12,AustinYFS13} in web
applications. Like \lweb, they have formalized and proved a
noninterference property (but not mechanized it).
Unlike \lweb that enforces IFC using the underlying \lio monad, 
Jacqueline at runtime explicitly keeps track of the secret and public views 
of sensitive values. While expressive, this approach can be expensive
in both space and time: results of computations on sensitive values 
have up to $1.75\times$ slower running times, and require more memory. 
Latencies for Django and Jacqueline 
are around 160ms for typical requests to their benchmark application.

The system most closely related to \lweb is
Hails~\cite{Giffin:2012:HPD:2387880.2387886,stefan17hails}, which aims
to enforce information flow-oriented policies in web
applications. Hails is also based on \lio, and is particularly
interested in confining third-party extensions (written in Safe
Haskell~\cite{Terei:2012:SH:2430532.2364524}). In Hails, individual
record fields can have policies determined by other data in the
database, as determined by a general Haskell function provided by the
programmer. Thus, Hails policies can encode \lweb policies, and more;
\eg data in one table can be used to determine labels for data in
another table. Evaluating the policy function during query processing
is potentially expensive. That said, according to their benchmarks,
the throughput of database writes of 
Hails is $2\times$ faster than Ruby Sinatra, comparable to Apache PHP, and $6\times$
slow than Java Jetty. They did not measure Hails' overhead, \eg by
measuring the performance difference with and without policy checks. 

There are several important differences between \lweb and
Hails. First, \lweb
builds on top of a mature, popular web framework (\yesod). Extracting
\lio into \lmonad makes it easy for \lweb to evolve as \yesod
evolves. As such, \lweb can benefit from \yesod's optimized code,
bugfixes, etc. Second, \lweb's @lsql@ query language is highly
expressive, whereas (as far as we can tell) Hails uses a simpler query
language targeting MongoDB where predicates can only depend on the document key. 
Third, there is no formal argument (and
little informal argument) that Hails' policy checks ensure a
high-level security property. The ability to run arbitrary code to
determine policies seems potentially risky (\eg if there are mutually
interacting policy functions), and there seems to be
nothing like our database invariants that are needed for noninterference. 
Our mechanized formalization proved important:
value-oriented policies (where one field's 
label depends on another field) were tricky to get right (per
\cref{sec:liquidhaskell-discussion}). 

Finally, IFDB~\cite{Schultz:2013:IDI:2465351.2465357} defines an
approach to integrating information flow tracking in an application
and a database. Like Hails and \lweb, the application tracks a current
``contamination level,'' like \lio's current label, that reflects data it
has read. In IFDB, one can specify per-row policies using secrecy and
integrity labels, but not policies per
field. Labels are stored as separate, per-row metadata, implemented by
changing the back-end DBMS. Declassification is permitted within
trusted code blocks. Performance overhead for HTTP request latencies
was similar to \lweb, at about 24\%. Compared to IFDB, \lweb
does not require any PSQL/database modifications; can support
per-field, updatable labels; and can treat existing fields as labels,
rather than requiring the establishment of a separate (often
redundant) field just for the label. IFDB also lacks a clear argument
for security, and has no formalization. Once again, we found such a
formalization particularly useful for revealing bugs.

\section{Conclusion}\label{sec:conclusion}

We presented \lweb, 
a information-flow security enforcement mechanism for Haskell web applications.
\lweb combines \yesod with \lmonad, a generalization of the \lio
library. \lweb performs label-based policy checks and protects 
database values with dynamic labels,
which can depend on the values stored in the database. 
We formalized \lweb (as \lwebcalc) and used Liquid Haskell to prove termination-insensitive 
noninterference. 
Our proof uncovered two noninterference violations in the implementation. 
We used \lweb to build the web site of the \emph{Build it,
  Break it, Fix it} security-oriented programming contest, and found
it could support rich policies and queries.
 Compared to manually checking security policies, 
\lweb impose a modest runtime overhead between \overheadnumbermin to \overheadnumber
but reduces the trusted code base to \tcbnumberbibifi of the
application code, and \tcbnumber  overall (when counting \lweb too).

\begin{acks}
We would like to thank Alejandro Russo and the anonymous reviewers for
helpful comments on a draft of this paper. This work was supported in
part by the \grantsponsor{GS100000001}{National Science Foundation}{http://dx.doi.org/10.13039/100000001} under grant \grantnum{GS100000001}{CNS-1801545} and
by \grantsponsor{GS100000185}{DARPA}{http://dx.doi.org/10.13039/100000185} under contract \grantnum{GS100000185}{FA8750-16-C-0022}.
\end{acks}

\bibliography{bib/refs.bib}

\end{document}